\documentclass[times,review,3p]{elsarticle}

\usepackage{lineno}

\usepackage[printwatermark]{xwatermark}
\usepackage{xcolor}
\usepackage{tikz}

\usepackage[bookmarks=false]{hyperref}

\journal{Computer Physics Communications}

\usepackage{amsmath}
\usepackage{graphicx}
\usepackage{booktabs}
\usepackage{threeparttable}
\usepackage{subcaption}
\usepackage{amssymb}
\usepackage{latexsym}
\usepackage{url}
\usepackage{xcolor}

\newsavebox\mybox
\savebox\mybox{\tikz[color=red,opacity=0.2]\node{PREPRINT};}
\newwatermark*[
  allpages,
  angle=45,
  scale=10,
  xpos=-50,
  ypos=50
]{\usebox\mybox}
\begin{document}

\begin{frontmatter}

\title{SOL-KiT - fully implicit code for kinetic simulation of parallel electron transport in the tokamak Scrape-Off Layer}%

\author[1]{S. Mijin\corref{mycorrespondingauthor}}
\author[1]{A. Antony}
\author[2]{F. Militello}
\author[1]{R.J. Kingham}
\address[1]{Blackett Lab., Plasma Physics Group, Imperial College, London SW7 2AZ, UK }
\address[2]{CCFE, Culham Science Centre, Abingdon, Oxon OX14 3DB, UK}
\cortext[mycorrespondingauthor]{Corresponding author}

\begin{abstract}
Here we present a new code for modelling electron kinetics in the tokamak Scrape-Off Layer (SOL). SOL-KiT (\textbf{S}crape-\textbf{O}ff \textbf{L}ayer \textbf{Ki}netic \textbf{T}ransport) is a fully implicit 1D code with kinetic (or fluid) electrons, fluid (or stationary) ions, and diffusive neutrals. The code is designed for fundamental exploration of non-local physics in the SOL and utilizes an arbitrary degree Legendre polynomial decomposition of the electron distribution function, treating both electron-ion and electron-atom collisions. We present a novel method for ensuring particle and energy conservation in inelastic and superelastic collisions, as well as the first full treatment of the logical boundary condition in the Legendre polynomial formalism. To our knowledge, SOL-KiT is the first fully implicit arbitrary degree harmonic kinetic code, offering a conservative and self-consistent approach to fluid-kinetic comparison with its integrated fluid electron mode. In this paper we give the model equations and their discretizations, as well as showing  the results of a number of verification/benchmarking simulations.
\end{abstract}


\end{frontmatter}

% Computer program descriptions should contain the following
% PROGRAM SUMMARY.

{\bf PROGRAM SUMMARY}
  %Delete as appropriate.

\begin{small}
\noindent
{\em Program Title: } SOL-KiT                              \\
{\em Licensing provisions: }                       GNU  GPLv3             \\
{\em Programming language: }               Fortran 90                    \\
{\em Nature of problem: }Fluid models of parallel transport in the Scrape-Off Layer (SOL) fail to account for the fact that the electron disctribution function is often far from a Maxwellian, and kinetic effects have been linked to discrepencies between experiment and fluid modelling[1]. A kinetic treatment of electrons in the SOL requires detailed accounting of collisional processes, especially those with neutral particles, as well as a proper implementation of the logical boundary condition at the material surface[2]. Furthermore, the ability to identify differences between fluid and kinetic modelling using self-consistent comparison is desirable.\\
{\em Solution method:} Electrons are modelled either as a fluid, or kinetically, maintaining self-consistency between models. All equations are solved using finite difference and the implicit Euler method, with fixed-point iteration. The kinetic approach is based on solving the Vlasov-Fokker-Planck-Boltzmann equation, decomposed in Legendre polynomials. Equations for the harmonics (or electron fluid equations) are solved alongside fluid equations for the ions, Amp\`{e}re-Maxwell's law for the electric field, as well as a diffusive-reactive Collisional-Radiative model for the evolution of hydrogenic atomic states. Each individual operator is built into a matrix, combined with all other operators, and the matrix equation arising from the Euler method is solved using the PETSc library, with MPI parallelization.\\
{\em Additional comments: }This article presents the physical and numerical outline of the code, and is accompanied by html documentation, a small test suite based on benchmarking runs presented below, as well as instructions and means for compiling and executing SOL-KiT. Special focus in the article is given to the novel numerical and model aspects in the greater context of the developed software.\\
  %Provide any additional comments here.
   \\

\end{small}

\section{Introduction}

The heat flow onto the plasma facing components of both present day and future magnetically confined fusion (MCF) devices  is of considerable importance \cite{Lipschultz2007, Pitts2011}, as it will greatly affect the lifetime of the material. This is true in both steady state operation and during transients (such as ELMs - Edge Localalized Modes). Understanding the heat flux in the Scrape-Off Layer (SOL) is thus of key importance for the design and operation of future fusion devices.

Classic fluid modelling of the parallel (to the magnetic field lines) energy transport in the edge region of MCF devices has relied on the fluid closure of Braginskii \cite{Braginskii1965}, or otherwise on various flux limiter approaches \cite{Fundamenski2005}. However, it is now well known that there exist discrepancies between experiments and the widely used fluid codes. These discrepencies can be, at least partly, attributed to the effect of non-local transport in the SOL \cite{Chankin2009}. In this paper, the term ``non-local'' is used to describe behaviour that stems from strong departure of the electron distribution function from a Maxwellian, in particular due to the fact that electron-ion mean-free paths in situations of interest are comparable to or greater than the temperature gradient scale lengths. 

In the following text we focus mainly on aspects of the divertor SOL. The main feature of the divertor configuration is that the location of the primary plasma-surface interaction is relatively far away from the hot core \cite{Stangeby2000} to specifically designed target plates. We distinguish between the ``upstream'', closer to the core, and the ``downstream'', where the plasma near the divertor targets is considerably cooler, and the ionization degree can be well below 100\%, rendering plasma-neutral interaction important. As such, a large temperature gradient is present, and plasma collisionality (measured with the electron-ion collision mean free path $\lambda_{ei}$) varies greatly along the magnetic field lines in the SOL. Of critical importance is the ratio of the mean free path to the temperature gradient scale length $L_{\nabla T} =(\nabla_{||}T/T)^{-1}$. Once the ratio $\lambda_{ei}/L_{\nabla T}$ is no longer much less than unity, the classical transport results are no longer valid \cite{Fundamenski2005}. However, another important concept in the understanding of energy transport in the SOL is that of the high energy heat-carrying electrons (HCE), which become marginally collisionless before the bulk of the distribution \cite{Chodura1990,Chankin2015}, and will remain so in most SOL situations. In other words, even if the ratio $\lambda_{ei}/L_{\nabla T}$ might still imply the correctness of classical transport coefficients, the HCE could be collisionless, and thus modify the transport by producing non-Maxwellian distribution functions. A further complication in the understanding of the SOL, as mentioned above, is the importance of electron-neutral interactions. This is especially true during detachment\cite{Krasheninnikov2017}, when the ionization degree drops considerably, and a neutral cloud is formed between the divertor targets and the upstream plasma. As this regime of operation offers better protection to the divertor plate materials, it becomes important to understand the interplay of kinetic/non-local effects already present in the SOL with detachment. 

\subsection{SOL kinetic modelling}

In order to properly capture non-local effects in the SOL it is necessary to treat the plasma using a kinetic approach. Broadly speaking, the two main approaches in the kinetic modelling of the SOL are the Particle-in-Cell (PIC) and finite difference methods solving the Vlasov-Fokker-Planck equation sometimes combined with the Boltzmann collision integral.

A representative PIC code for SOL simulations is BIT1\cite{Tskhakaya2008,Tskhakaya2012}, used in a variety of simulation scenarios corresponding to present day machine conditions. The PIC method is highly parallelizable, and naturally accommodates the addition of many different collision types (e.g. tungsten impurities\cite{Tskhakaya2013,Kirschner2018}). While detailed simulations covering many aspects of SOL transport are possible, two issues make PIC codes complicated to operate.. These are the usually long run times (compared to finite difference methods and fluid codes), and the fact that the number of particles simulated in a PIC code can never approach reality, requiring smoothing techniques\cite{Tskhakaya2010}, and potentially not resolving the high energy tails of distributions with enough accuracy.

Finite difference methods do not suffer from the noise problems of PIC codes, as they solve for the distribution function directly. A number of finite difference codes with different approaches have been utilized in the modeling of the SOL, with a few examples mentioned here. An early example of a completely kinetic code (treating every species kinetically) was the ALLA code\cite{Batishchev1999}. Another code, utilizing a similar method to what is presented in this paper, albeit with an explicit algorithm, is the FPI code\cite{Abou-Assaleh1992,Abou-Assaleh1994,Allais2005}, where electrons are treated kinetically while others species are stationary. More recently, the code KIPP\cite{Chankin2015,Zhao2017,Chankin2018}, with kinetic electrons, has been coupled with the 2D fluid code SOLPS, providing the latter with kinetically calculated transport coefficients. 

\subsection{Motivation to develop SOL-KiT}

Due to the great mass difference between electrons and other species within a hydrogen plasma, electrons mainly suffer pitch-angle scattering collisions when colliding with those heavier particles. Eigenfunctions of such collision operators are spherical harmonics, and an expansion of the electron distribution function in spherical harmonics becomes natural\cite{Shkarofsky}. This approach has been used in the modeling of Scrape-Off-Layer transport to a limited extent\cite{Abou-Assaleh1992,Abou-Assaleh1994,Allais2005}, but has been used both in codes dealing with laser-plasma interactions (KALOS\cite{Bell2006}/OSHUN\cite{Tzoufras2011}, and IMPACT\cite{Kingham2004}), as well as electron swarm transport models\cite{Makabe,Kumar1980}. The expansion has been used to efficiently model both Coulomb and electron-neutral collisions, and has proven itself to be a powerful tool in treating plasmas of various collisionality. With this in mind, a marriage of the Vlasov-Fokker-Planck (VFP) approach in laser plasmas and the Boltzmann approach of electron swarms in neutral gases seems to be potentially a highly applicable model for the Scrape-Off Layer plasma, where collisionality changes, and neutral-plasma interactions carry a great deal of importance. 

As is typical with solutions of differential equations, the boundary conditions tend to define the system behaviour and dictate the approach in the numerical solution. In modeling the SOL, it becomes necessary to incorporate the effect of the plasma sheath formed at the boundary, i.e. at the divertor target. While the traditional approach of Procassini et al.\cite{Procassini1990} has been used in many kinetic codes, when utilizing the spherical harmonic expansion it becomes necessary to formulate the well known boundary condition in terms of the expansion basis. We present this formulation, and give its implementation in SOL-KiT.

The divertor target plate acts as a sink of particles, and as such generates flows towards the target. Since the divertor boundary condition is formulated in the lab frame, it is then necessary to treat the ion flow in the lab frame as well. As a consequence, the electron-ion collision operator must be extended to account for the moving ions. This is a different strategy for incorporating ion motion in the electron VFP equation, than used elsewhere.  There the Vlasov terms are transformed instead into the local rest frame of the ions\cite{Epperlein1988}.  We present a simplified treatment of this lab frame operator in the next section, together with fluid ion equations. In order to be able to resolve the low velocity cells for higher harmonics (avoiding the CFL condition) it then becomes necessary to treat the electron kinetic equation implicitly.

Another consequence of the divertor boundary conditions is that it also acts as a source of neutral particles. The inclusion of electron-neutral collisions on a nonuniform velocity grid poses particle and energy conservation problems. We present a method of mapping inelastic collisions on a nonuniform velocity grid which conserves both energy and particles, as well as obeying a numerically consistant detailed balance condition when calculating superelastic collision cross-sections. In order to self-consistently model the interaction of atomic states and the electrons, we include a diffusive-reactive Collisional Radiative model\cite{Capitelli2016} for the evolution of hydrogenic atomic states.

Finally, in order to provide a modular, self-consistent one-to-one comparison of kinetic and fluid modelling, the code also includes fluid equations for the electrons, which can be solved instead of the kinetic model, while making sure that all of the physical processes are the same, and that the atomic data is used consistently between the two models.

To the authors' knowledge, this is the first fully implicit arbitrary Legendre polynomial/Spherical Harmonic code that has an inbuilt sheath boundary condition, inelastic electron-neutral collisions, and a self-consistent fluid mode for clean comparisons. In the following sections the equations of SOL-KiT and their numerical implementation will be presented, starting with the analytical aspects of the model in section 2, before moving on to the model's numerical implementation in section 3. Finally, details of performed benchmarking runs will be given in section 4. We discuss the various aspects of the code in section 5. 

\section{Physical model}

In this section we will introduce the equations being solved in the SOL-KiT model, giving a condensed overview of the physics before expanding on individual operators. 

\begin{figure*}[htbp]
\centering
\includegraphics[width=0.75\textwidth]{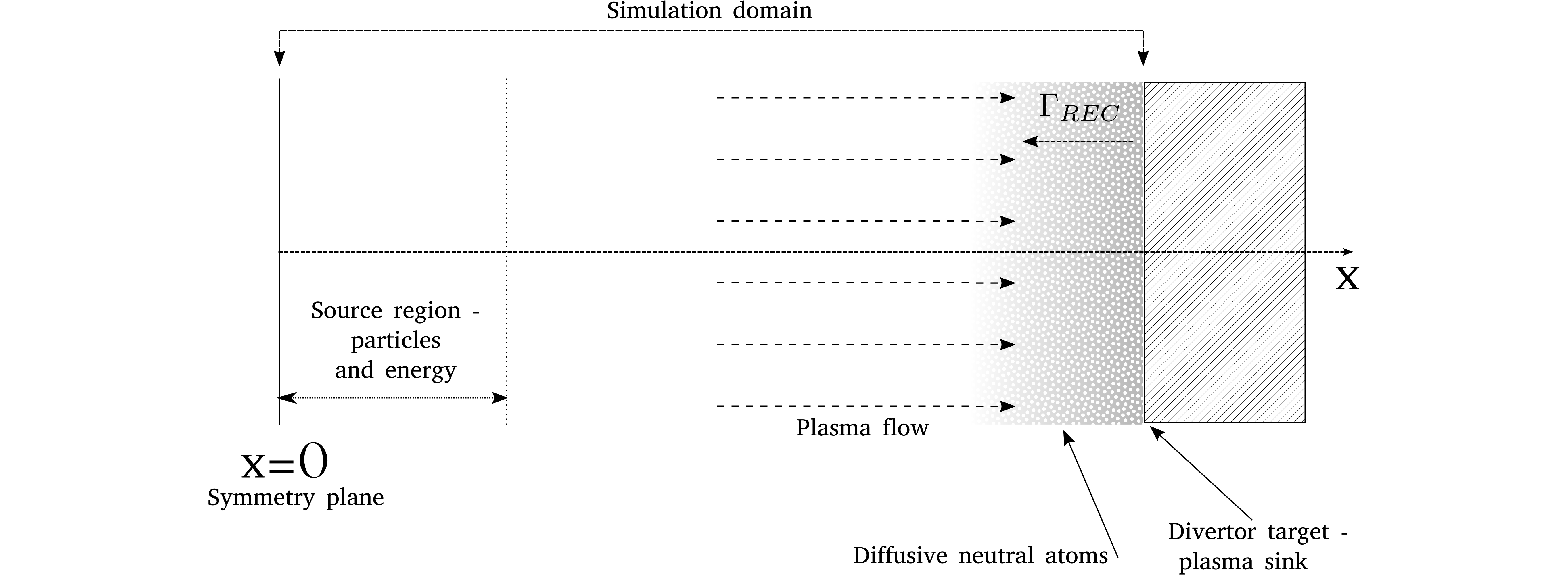}
\caption{The SOL simulation domain (not to scale): the $x$-axis is the principle axis of the system, with $x=0$ being the upstream symmetry plane; the right boundary of the system is at the divertor target, which acts as a sink for the plasma, as well as a source of neutrals via the recycling flux $\Gamma_{REC}$}
\label{fig:domain}
\end{figure*}

While the code is capable of handling both fixed and periodic boundary conditions, since the most involved cases utilize the Scrape-Off Layer domain, we start with describing it. The domain is 1D, and is assumed to be along a (straightened) field line. The field line is taken as the $x$-axis, around which the domain is symmetric. The point $x=0$ is taken to be the symmetry plane, representing the ''upstream'' of the SOL. A sketch of the simplified SOL domain is given in Figure \ref{fig:domain}.

The equations solved by SOL-KiT are the following:

\begin{itemize}
\item Electron equations - either fluid (density, parallel velocity, temperature) or kinetic
\item Ion fluid equations - density and parallel velocity (assuming either $T_i = 0$ or $T_i = T_e$)
\item Diffusive-reactive Collisional Radiative Model for atomic states
\item Amp\`{e}re-Maxwell law for the evolution of the electric field
\end{itemize}

For the electrons we either solve the kinetic equation, which is the main mode of the code, or we can solve local fluid equations, obtained by taking moments of the kinetic model, ensuring maximum correspondance between the kinetic and fluid modes. This in turn allows for easy comparison between the fluid and kinetic model, further highlighting kinetic effects. The 1D kinetic equation solved for the electrons is the Vlasov-Fokker-Planck-Boltzmann equation, given by 

\begin{equation}
\frac{\partial f(x,\vec{v},t)}{\partial t} + v_x \frac{\partial f(x,\vec{v},t)}{\partial x}  -\frac{e}{m_e} E \frac{\partial f(x,\vec{v},t)}{\partial v_x} = C[f,...],
\label{eq:e-kin}
\end{equation} 
where $E$ is the electric field (assuming azimuthal symmetry and straightening out the magnetic field we ignore magnetic effects). The RHS contains all of the collision and source operators. Details on collision operators and the Legendre polynomial decomposition of the electron distribution function are given in Section 2.1. In the electron fluid mode, the continuity, momentum, and temperature equations are solved instead. 

The continuity equation is given by

\begin{equation}
\frac{\partial n_e}{\partial t} + \frac{\partial (n_e u_e)}{\partial x} = S,
\end{equation}

while the momentum equation is
\begin{equation}
\frac{\partial u_e}{\partial t} = - u_e  \frac{\partial u_e}{\partial x}  - \frac{e}{m_e} E + \frac{R_{ei}+R_{en}}{m_e n_e} - \frac{S}{n_e}u_e - \frac{1}{m_e n_e}\frac{\partial( n_e k T_e)}{\partial x},
\label{eq:el-mom}
\end{equation}
where $n_e$,$u_e$, and $T_e$ are the electron density, flow velocity, and temperature, respectively. $S=S_{ion} + S_{rec}$ is the particle source including ionization and recombination, and $R_{ei}=R_T+R_u$ is the classical Braginskii\cite{Braginskii1965} friction

\begin{equation}
R_u = - \frac{m_e n_e}{\tau_e} 0.51 (u_e-u_i),
\end{equation}

\begin{equation}
R_T = - 0.71 n_e \frac{\partial (kT_e)}{\partial x},
\end{equation}
where the $\tau_e$ is the Braginskii collision time\cite{Braginskii1965}. $R_{en}$ is the total friction from all electron-neutral collisions, assuming a slowly (compared to electron thermal speed) drifting Maxwellian distribution for electrons.  

The electron temperature equation is

\begin{equation}
\frac{\partial kT_e}{\partial t} = - u_e  \frac{\partial kT_e}{\partial x}  + \frac{2}{3} \left[\frac{Q}{n_e} - kT_e\frac{\partial u_e}{\partial x} - \frac{1}{n_e}\frac{\partial q_e}{\partial x} - \frac{S}{n_e}\left(\frac{3}{2}kT_e-\frac{m_eu_e^2}{2}\right) - \frac{u_e (R_{ei}+R_{en})}{m_en_e}\right],
\label{eq:el-temp}
\end{equation}
where $q_e=q_T+q_u$ is again the classical Braginskii\cite{Braginskii1965} heat flux

\begin{equation}
q_T = - \kappa_e \frac{\partial (kT_e)}{\partial x},
\label{eq:SH-q}
\end{equation}

\begin{equation}
q_u = 0.71 n_e k T_e (u_e - u_i),
\end{equation}
where $\kappa$ is taken to be either the Spitzer-H\"{a}rm result or the Lorentz result \cite{Epperlein1984,Epperlein1986}, depending on whether we treat electron-electron momentum transfer collisions. 
$Q=Q_{ext}+Q_{en}$ is the combination of the external heating energy source (see 2.1.4.), as well as any inelastic collision energy transfer between the electrons and the atoms.

The ion equations are analogous, with a few differences. Firstly, the ion continuity equation can be solved, or quasi-neutrality can be enforced artificially by setting $Zn_i=n_e$. Secondly, there is no ion temperature equation in the current version of the code. Instead, ion temperature is set to either zero, or to the electron temperature. The equations are 

 \begin{equation}
\frac{\partial n_i}{\partial t} + \frac{\partial (n_i u_i)}{\partial x} = S ,
\end{equation}

\begin{equation}
\frac{\partial u_i}{\partial t} = - u_i  \frac{\partial u_i}{\partial x}  + \frac{Ze}{m_i} E + \frac{R_{ie}+R_{CX}}{m_i n_i} - \frac{S}{n_i}u_i - \frac{1}{m_i n_i}\frac{\partial( n_i k T_i)}{\partial x},
\label{eq:ion-mom}
\end{equation}
where $R_{ie}$ is obtained by requiring total momentum in electron-ion collisions to be conserved, ie. $R_{ie}=-R_{ei}$. $R_{CX}$ is a simple charge exchange friction term, given by 

\begin{equation}
R_{CX} = - n_i m_i u_i |u_i|\sum_b n_b \sigma_{CX,b},
\end{equation} 
where the sum is over neutral atomic states, and both the ions and neutrals are approximated as cold, with the simplified constant hydrogenic charge exchange cross sections given by approximate low energy values obtained from Janev\cite{Janev2003}

$$\sigma_{CX,1} = 3\times 10^{-19}m^2, \quad \sigma_{CX,2} = 2^4 \times 10^{-19}m^2, \quad \sigma_{CX,3} = 3^4 \times 7 \times 10^{-20}m^2, \quad \sigma_{CX,b \geq 4} = b^4 \times 6 \times 10^{-20}m^2.$$

To calculate the electric field, we solve Amp\`{e}re-Maxwell's law, which contains only the displacement current

\begin{equation}
\frac{\partial E}{\partial t} = - \frac{1}{\epsilon_0} (j_e + Ze n_i u_i),
\end{equation}
where $j_e$ is either given as a moment of the electron distribution function, or simply as $j_e = - e n_e u_e$ in the electron fluid case.

Finally, the atomic state distribution of the neutrals must be tracked. This is done using a diffusive-reactive Collisional Radiative model (CRM) to obtain the evolution of the neutral state densities $n_b$ (where $b$ here denotes the principal quantum number of the state)

\begin{align}
\frac{\partial n_b}{\partial t} & \notag =  \frac{\partial}{\partial x}\left(D_b \frac{\partial  n_b}{\partial x}\right) +  \sum_{b'<b} \left[  K_{b' \rightarrow b}^e n_{b'}-A_{b \rightarrow b'}n_b - K_{b \rightarrow b'}^e n_b \right] \\& + \sum_{b'>b} \left[K_{b' \rightarrow b}^e n_{b'} + A_{b' \rightarrow b} n_{b'} - K_{b \rightarrow b'}^e n_b\right] - K_{b}^{ion} n_b +  \alpha_b n_e^2n_i + \beta_b n_e n_i,
\label{eq:CRM}
\end{align}
where the ionization and (de-)excitation rates $K$, as well as three-body recombination rates $\alpha$ are calculated using moments of the distribution function (see 2.1.3.). The inelastic cross-sections and radiative de-excitation/recombination rates $A$ and $\beta$ are all taken from Janev\cite{Janev2003} and NIST\cite{NIST}. Radiative de-excitation is included only up to state number $b=20$, due to lack of available data. Since higher excited states are primarily collisionally coupled in most situations of interest this should not cause significant discrepancies.

The classical 1D diffusion coefficient is simply

\begin{equation}
D_b = \frac{v_{tn}}{2[(n_i+n_1)\sigma_{el} + \sigma_{CX,b}n_i]}
\end{equation}
where $v_{tn}$ is the thermal speed of neutrals, $\sigma_{el}$ is the elastic collisions cross-section (see electron-neutral elastic collision operator in 2.1.3.), and $n_1$ is the ground state density. When charge exchange is used, it is the dominant term in the diffusion coefficient, but the elastic collision diffusion is included for cases when charge exchange is turned off. Since gas temperature and elastic cross-section are free parameters in SOL-KiT, this operator can be tuned. Ideally, however, a self-consistent neutral dynamics model should be implemented, and this is a planned extension of the model.

The boundary condition at the divertor target is the logical boundary condition \cite{Procassini1990} when electrons are treated kinetically, while the ions are always assumed to reach sound speed (as per the Bohm criterion). Details of the kinetic boundary condition are given below. In both the fluid and kinetic model, the flow into the sheath is ambipolar $\Gamma_e = \Gamma_i$.  When electrons are treated as a fluid the sheath heat transmission coefficient\cite{Stangeby2000} in $q_{sh}=\gamma kT_e \Gamma_e$ is taken to be $\gamma_e=2 - 0.5\ln (2\pi(1+T_i/T_e)m_e/m_i)$. Finally, the atomic neutrals are recycled with a recyling flux $\Gamma_{REC}=-R\Gamma_i$, where $R \leq 1$ is the recycling coefficient. This is simply imposed by setting $D_1 \partial  n_1/\partial x$ in (\ref{eq:CRM}) at the boundary to $\Gamma_{REC}$, whereas it would otherwise be zero.

\subsection{Electron kinetic equation in Legendre formalism}

Spherical harmonics are an orthonormal basis set in the solid angle space $ \left( \theta,\varphi \right) \in \left[ 0,\pi \right) \times  \left[ 0, 2 \pi  \right)$, and can be written as a suitably normalized product of associated Legendre polynomials $P_l^m(\cos \theta)$ and the complex phase $e^{im\varphi}$. The spherical harmonic convention used in SOL-KiT is same as the ones in KALOS\cite{Bell2006}/OSHUN\cite{Tzoufras2011}. As such, the traditional Cartesian coordinate system for velocity $(v_x,v_y,v_z)$ will be rotated so that the angles $\theta$ and $\varphi$ are defined as part of a spherical coordinate system:

\begin{equation}
v = \sqrt{v_x^2 + v_y^2 + v_z^2}, \quad \theta = \arccos(v_x/v), \quad \varphi = \arctan(v_z/v_y).
\end{equation}

If then we decompose the distribution function $f(v,\theta,\phi)$ in spherical harmonics we get:

\begin{equation}
f(v,\theta,\varphi) = \sum_{l=0}^{\infty}{ \sum_{m=-l}^{l} {f_l^m(v) P_l^{|m|}(\cos\theta)e^{im\varphi}}}
\end{equation} 
where the complex expansion coefficients $f_l^m$ satisfy $(f_l^m)^* = f_l^{-m}$. This allows writing the kinetic equation for electrons as a set of equations for the amplitudes $f_l^m$, whose physical significance becomes obvious when moments of the distribution function are expressed using the expansion. If $\phi$ is a scalar function of $v$ then

\begin{equation}
\int \phi f(\vec{v})d\vec{v} = 4 \pi \int_0^{\infty} \phi f_0^0(v) v^2 dv
\end{equation} 
while if $\vec{a}$ is a vector function of $v$

\begin{equation}
\int \vec{a} f(\vec{v})d\vec{v} = \frac{4 \pi}{3} \int_0^{\infty} ||a|| \begin{pmatrix}f_1^0\\2Re(f_1^1)\\-2Im(f_1^1)\end{pmatrix} v ^2 dv
\end{equation} 
and similarly for higher order tensors\cite{Shkarofsky,Tzoufras2011}. As can be seen from this, not only does the spherical harmonic expansion have useful properties in relation to collisions, it also provides a physically meaningful decomposition for evaluating transport quantities.

Since the current model is 1D, and azimuthal symmetry around the $x$-axis is assumed, we do not treat magnetic field effects, and the spherical harmonic decomposition reduces to a Legendre polynomial decomposition  ($m=0$ always). Accordingly, in the rest of this paper, harmonics will be labeled only by their $l$-number, i.e. $f_l(v)$. 

Following the Legendre decomposition of the distribution function as outlined above, the 1D kinetic equation can be written as a set of equations

\begin{equation}
\frac{\partial f_l(x,v,t)}{\partial t} = A_l + E_l + C_l,
\label{eq:e-kin-l}
\end{equation} 
where now all of the operators are moved to the RHS and are functions of $l$ in addition to whatever arguments they naturally have.  These will be examined in detail below.

\subsubsection{Vlasov terms}

The terms on the LHS of equation (\ref{eq:e-kin}) are usually referred to as the Vlasov terms. The two Vlasov terms in equation (\ref{eq:e-kin-l}) are the spatial advection term $A_l$ (corresponding to second LHS term in (\ref{eq:e-kin})), and the velocity space advection term due to the electric field in the $x$-direction $E_l$ (corresponding to third LHS term in (\ref{eq:e-kin})). 

Firstly, the spatial advection term (advection in the $x$-direction), for a given harmonic $l$ is

\begin{equation}
A_l = -\frac{l}{2l-1}v\frac{\partial f_{l-1}}{\partial x} - \frac{l+1}{2l+3}v\frac{\partial f_{l+1}}{\partial x}.
\label{eq:A-l}
\end{equation}
Spatial advection couples harmonics with different $l$ numbers. The physical significance of this coupling is most easily seen in the coupling between $f_0$ and $f_1$. The moments of $f_0$ are the density and total energy, while $f_1$ is associated with flows (of particles and energy). Thus gradients in $f_0$ (density and temperature) drive advection of $f_1$ (flows), and vice-versa.

The velocity space advection term due to the electric field couples harmonics as well, albeit through velocity space gradients in $f_l$. As only the $x$ component of the electric field is treated, in the remainder of the paper it will simply be written as $E$ for brevity. The velocity space advection operator is given by\cite{Bell2006}

\begin{equation}
E_l =  \frac{e}{m}E \left[\frac{l}{2l-1}G_{l-1} + \frac{l+1}{2l+3}H_{l+1}\right]
\label{eq:E-l}
\end{equation}
where

\begin{equation}
G_l(v) = v^l \frac{\partial v^{-l} f_l}{\partial v},
\label{eq:G-l}
\end{equation}

\begin{equation}
H_l(v) = \frac{1}{v^{l+1}} \frac{\partial v^{l+1} f_l}{\partial v}.
\label{eq:H-l}
\end{equation}
As is evident from equation (\ref{eq:E-l}), the electric field couples harmonics through the $G_l$ and $H_l$ functions, which contain velocity space gradients of the coupled harmonics.

Thus, Vlasov terms provide coupling of harmonics through either spatial gradients or the electric field.

\subsubsection{Coulomb collision terms}

 Let us consider the effect of Coulomb collisions on the distribution function $f$ of particles of mass $m$ and charge $q=ze$ colliding with particles of mass $M=\mu m$ and charge $Q=Ze$ with distribution $F$. We follow the formalism of Shakorfsky et al.\cite{Shkarofsky}, starting from the Rosenbluth coefficient formulation of the Fokker-Planck collision operator for Coulomb collisions

\begin{equation}
\frac{1}{\Gamma_{zZ}} \frac{\delta f}{\delta t} = \frac{4 \pi}{\mu} Ff + \frac{\mu - 1}{\mu + 1} \nabla \mathcal{H}(F)\cdot \nabla f + \frac{\nabla \nabla \mathcal{G}(F) : \nabla \nabla f}{2},
\label{eq:RosenbluthFP}
\end{equation}
where $\nabla = \partial / \partial \vec{v}$ and $\Gamma_{zZ} = (zZe^2)^2\ln\Lambda/(4\pi(m\epsilon_0)^2)$, with $\ln\Lambda$ denoting the Coulomb logarithm. The Rosenbluth drag and diffusion coefficients are respectively $\mathcal{H}$ and $\mathcal{G}$. We separate the distribution functions into their isotropic and anisotropic components $F = F_0 + F_a$, $f = f_0 + f_a$. The key assumption going forward is that the anisotropic component is small compared to the isotropic one, so that it becomes possible to linearize equation (\ref{eq:RosenbluthFP}). Expanding the distribution function and the Rosenbluth coefficients in harmonics and using the integrals\cite{Shkarofsky}

\begin{equation}
I_j(F_l)=\frac{4\pi}{v^j}\int_0^{v} F_l(u)u^{j+2}du, \quad J_j(F_l)=\frac{4\pi}{v^j}\int_v^{\infty} F_l(u)u^{j+2}du,
\end{equation} 
one can derive the expressions for the harmonic components of the Fokker-Planck collision integral for all $l$. For $l=0$ this is

\begin{equation}
\frac{1}{\Gamma_{zZ}} \frac{\delta f_0}{\delta t} = \frac{1}{3v^2} \frac{\partial}{\partial v} \left[\frac{3}{\mu}f_0 I_0(F_0) + v\left(I_2(F_0) + J_{-1}(F_0)\right)\frac{\partial f_0}{\partial v}\right].
\label{eq:FP-l0}
\end{equation}

For $l>0$ the following is obtained\cite{Tzoufras2011}

\begin{align}
\frac{1}{\Gamma_{zZ}} \frac{\partial f_l}{\partial t} & \notag= \frac{4 \pi}{\mu} \left[F_0 f_l + f_0 F_l\right] \\
										       & \notag - \frac{(\mu - 1)}{\mu v^2} \left\{ \frac{\partial f_0}{\partial v} \left[\frac{l + 1}{2l+1} I_l(F_l) - \frac{l}{2l+1}J_{-1-l}(F_l)\right] + I_0(F_0)\frac{\partial f_l}{\partial v} \right\} \\
& \notag+ \frac{I_2(F_0) + J_{-1}(F_0)}{3v} \frac{\partial^2 f_l}{\partial v^2} + \frac{-I_2(F_0) + 2J_{-1}(F_0) + 3I_0(F_0)}{3v^2} \frac{\partial f_l}{\partial v} \\
& \notag -\frac{l(l+1)}{2} \times \frac{-I_2(F_0) + 2J_{-1}(F_0) + 3I_0(F_0)}{3v^3} f_l \\
& \notag +\frac{1}{2v}\frac{\partial^2 f_0}{\partial v^2}\left[C_1 I_{l+2}(F_l) + C_1J_{-1-l}(F_l) + C_2 I_l(F_l) + C_2J_{1-l}(F_l)\right] \\
&+\frac{1}{v^2} \frac{\partial f_0}{\partial v} \left[C_3 I_{l+2}(F_l) + C_4J_{-1-l}(F_l) + C_5 I_l(F_l) + C_6J_{1-l}(F_l)\right],
\label{eq:FP-l>0}
\end{align}
where the $C$ coefficients are functions of $l$

\begin{align*}
C_1 &= \frac{(l+1)(l+2)}{(2l+1)(2l+3)}, \quad C_2 = -\frac{(l-1)l}{(2l+1)(2l-1)}, \quad C_3 = -\frac{(l+1)l/2 + l + 1}{(2l+1)(2l+3)}, \\
C_4 &= \frac{-(l+1)l/2 + l + 2}{(2l+1)(2l+3)}, \quad C_5 = \frac{(l+1)l/2 + l - 1}{(2l+1)(2l-1)}, \quad C_6 = -\frac{(l+1)l/2  -  l}{(2l+1)(2l-1)}.
\end{align*} 

For electron-electron collisions $\mu = 1$. The effect of e-e collisions on the isotropic part of the distribution function is given by

\begin{equation}
\frac{1}{\Gamma_{ee}} \left(\frac{\delta f_0}{\delta t} \right)_{e-e}= \frac{1}{v^2} \frac{\partial}{\partial v} \left[C(f_0)f_0 + D(f_0)\frac{\partial f_0}{\partial v}\right],
\label{eq:FP-l0-ee}
\end{equation}
where the drag and diffusion coefficients are defined as

\begin{equation}
C(f_0) = 4 \pi \int_0^{v} f_0(u)u^2du,
\end{equation} 

\begin{equation}
D(f_0) = 4 \pi \int_0^{v} u^2 \left[ \int_u^{\infty} f_0(u')u'du'\right]du.
\end{equation} 
Note that $D$ is not in the form one would expect from equation (\ref{eq:FP-l0}). Instead, it is written in an analytically equivalent form (see section 3.5.). The electron-electron collision operator for $l=0$ is important for the proper relaxation of the electron distribution function to a Maxwellian.

For higher harmonics, from (\ref{eq:FP-l>0}) we get

\begin{multline}
\begin{split}
\frac{1}{\Gamma_{ee}}\left(\frac{\delta f_l}{\delta t}\right)_{e-e} &= 8 \pi f_0 f_l +\frac{ I_2(f_0)+J_{-1}(f_0)}{3v}\frac{\partial ^2 f_l}{\partial v^2} \\&+ \frac{-I_2(f_0) + 2J_{-1}(f_0) + 3I_0(f_0)}{3v^2} \frac{\partial f_l}{\partial v} \\&- \frac{l(l+1)}{2} 
\times \frac{-I_2(f_0) + 2J_{-1}(f_0)+ 3I_0(f_0)}{3v^3} f_l \\& +
\frac{1}{2v}\frac{\partial^2 f_0}{\partial v^2} \left[C_1I_{l+2}(f_l)+C_1J_{-l-1}(f_l)+C_2I_l(f_l)+C_2J_{1-l}(f_l)\right] \\ 
&+ \frac{1}{v^2}\frac{\partial f_0}{\partial v} \left[ C_3I_{l+2}(f_l) + C_4J_{-l-1}(f_l) + C_5I_l(f_l) + C_6J_{1-l}(f_l)\right].
\end{split}
\label{FP-l>0-ee}
\end{multline} 

As the ions are either assumed cold or with the same temperature as the electrons, in the current version of SOL-KiT the effect of electron-ion collisions on $f_0$ is not included.

For higher harmonics we distinguish two cases of electron-ion collisions. The first is the classical stationary ion case, where $F_0 = n_i \delta(v)/(4 \pi v^2)$. It can easily be shown that equation (\ref{eq:FP-l>0}) reduces to the following eigenfunction form

\begin{equation}
\left(\frac{\delta f_l}{\delta t}\right)_{e-i} =  -\frac{l(l+1)}{2} \frac{\Gamma_{ei} n_i}{v^3} f_l.
\label{eq:FP-l>0-ei}
\end{equation}
Here it can be seen that this is purely angular scattering, which dampens higher harmonics, helping us truncate the expansion.

However, when ions are not stationary, but are moving at some velocity much smaller than the electron thermal velocity (the case we expect in the SOL), it is necessary to modify the electron-ion collision operator. This is done by first getting rid of all the terms proportional to the inverse mass ratio in equation (\ref{eq:FP-l>0}). To calculate the $I$ and $J$ integrals for a cold ion stream we let the ion distribution function be 

\begin{equation}
F(\vec v) = n_i \delta(\vec v - \vec{u_i}),
\end{equation} 
Recasting the Dirac delta into spherical coordinates assuming azimuthal symmetry and expanding in Legendre polynomials gives us

\begin{equation}
F_l(v) = \frac{n_i(2l+1)}{4\pi v^2} \delta(v - u_i).
\end{equation} 
Substituting these harmonics into the equations for the $I$ and $J$ integrals gives us

\begin{equation}
I_j(F_l)= (2l+1) n_i \frac{u_i^j}{v^j} \Theta(v-u_i),
\label{Ij-ci}
\end{equation} 

\begin{equation}
J_j(F_l)= (2l+1) n_i \frac{u_i^j}{v^j} \Theta(u_i - v),
\label{Jj-ci}
\end{equation} 
where $\Theta$ denotes the Heaviside step function.
It is now trivial to see that all but the $I_0$ integrals vanish when $u_i = 0$, recovering the stationary ion collision integral. It can also be easily shown that for small enough $u_i$, the collision integral for $f_1$ reduces to\cite{Shkarofsky}

\begin{equation}
\frac{1}{\Gamma_{zZ}} \frac{\partial f_1}{\partial t} = - \frac{n_i}{v^3}\left(f_1 + u_i \frac{\partial f_0}{\partial v}\right).
\end{equation}
If $f_0$ is taken to be Maxwellian, this gives a stationary solution to the $f_1$ collision operator to be that $f_1$ which yields a slowly drifting Maxwellian with drift velocity $u_i$.

Taking a closer look at the obtained collision integral, we see that electrons slower than the ions are being carried around by them, as one would expect. This modifies all harmonics for small $v$, so we lose the convenient property of clean truncation of the distribution function by electron-ion collisions. However, in realistic simulation cases, this happens only for a small handful of velocity space cells, while higher harmonics are normally dampened in the rest of  velocity space. This is to be expected, as our electrons see an ion Dirac delta and are collisionally driven towards it. 
\subsubsection{Boltzmann collision terms}

  We start from the general form of the Boltzmann collision integral for the effect of collisions on the distribution function of species $s$ colliding with species $s'$
 
 \begin{equation}
C[f_s,f_{s'}] (v)= \int d\vec{v_2} d\Omega |\vec{v}-\vec{v_2}|\sigma(|\vec{v}-\vec{v_2}|,\Omega)[f_s(\vec{v'})f_{s'}(\vec{v_2'})-f_s(\vec{v})f_{s'}(\vec{v_2})],
\label{eq:Boltz}
\end{equation} 
where primed velocities denote values before a collision, and $\sigma$ is the appropriate differential cross-section. The following results are all derived under the assumption of a small mass ratio and stationary (slow compared to the electrons) neutral particles (atoms)\cite{Shkarofsky,Makabe, Kumar1980}. Furthermore, all differential cross-sections are assumed to be azimuthally symmetric, i.e. are only a function of energy/velocity of the impacting particle (electron), and the deflection angle (here $\chi$). This just means that the cross-sections do not depend on the orientation of the neutral particle.

\paragraph{Elastic electron-neutral collisions}The first term in the small mass expansion of the electron-neutral elastic collision integral cancels for $l=0$. It is therefore necessary to include higher order effects of collisional energy transfer. This way, one can allow for long term relaxation of the electron distribution function to a Maxwellian with temperature equal to the gas temperature $T_g$. If $M$ is the neutral mass, and $n_b$ the gas density, the collision integral takes the form

\begin{equation}
\left(\frac{\delta f_0}{\delta t}\right)_{e-n_b}= \frac{m_e}{M+m_e}\frac{1}{v^2}\frac{\partial}{\partial v}\left[n_bv^4 \left( \int d\Omega (1 - \cos\chi) \sigma_b^{el}(\chi,v)\right) \left(f_0 + \frac{kT_g}{m_ev} \frac{\partial f_0}{\partial v} \right) \right]
\label{eq:en-el-0}
\end{equation}

As is the case with electron-ion Coulomb collisions, due to a great mass difference, the $l>0$ integral is considerably simplified, and can be written as

\begin{equation}
\left(\frac{\delta f_{l>0}}{\delta t}\right)_{e-n_b} = - n_b v \left[\int d\Omega (1-P_l(\cos(\chi)))\sigma_b^{el}(\chi,v) \right] f_l.
\label{eq:en-el-l}
\end{equation} 
where $P_l$ is simply the $l$-th Legendre polynomial. 

While implemented and tested in the current version of SOL-KiT, these two processes are rarely used, because of the lack of proper elastic collision cross-section data. Currently the cross-section for elastic collisions of an electron with a hydrogen atom in a given state is obtained using the classical expression for orbit size. Namely, this gives for the integral elastic collision cross-section of electrons with hydrogen atoms in the $b$-th state

\begin{equation}
\sigma_b^{el,TOT} = \pi a_0^2 b^4
\label{eq:sigma_el}
\end{equation}
where $a_0$ is the Bohr radius.

\paragraph{Inelastic electron-neutral collisions} We start with inelastic collisions where the total number of particles of each species is conserved (e.g. excitation). A standard procedure exists\cite{Shkarofsky,Makabe} for an inelastic collision for which the pre-collision and post-collision velocities are related as

\begin{equation}
\frac{m_ev'^2}{2} = \frac{m_ev^2}{2} + \epsilon
\label{eq:inel-rel}
\end{equation} 
where $\epsilon$ is the inelastic energy loss. Defining $\alpha = v'/v = (1+2\epsilon/mv^2)^{1/2}$, one can write the collision integral as

\begin{equation}
\left(\frac{\delta f_l}{\delta t} \right)^{ex}_{b\rightarrow b'} = -  n_b v\left[\sigma^{TOT}_{b\rightarrow b'}(v) f_l(v) - f_l(\alpha v) \alpha^2 \left(\sigma^{TOT}_{b\rightarrow b'}(\alpha v) -\sigma^{(l)}_{b\rightarrow b'}(\alpha v)\right) \right].
\label{eq:en-ex}
\end{equation}
Here $\sigma^{TOT}=\int d\Omega \sigma(\chi,v)$ is the integral cross section, while

\begin{equation*}
\sigma^{(l)}(v) =\int d\Omega (1 - P_l(\cos\chi)) \sigma(\chi,v).
\end{equation*} 
 Using equation (\ref{eq:en-ex}) for $l=0$ one can easilly show that particle number is conserved.

On the other hand, collisional processes such as ionization do not conserve the total number of particles. The main difficulty in treating ionization is the fact that it is, ultimately, a 3-body process, and ideally one would like to know the triply differential cross section for such a process. However, such an approach is not only complicated, but cross-section data (to the author's knowledge) are not systematically available. Because of this, the approach taken here will be the simplest possible\cite{Batishchev1999}, where all electrons produced in ionization are put in the lowest velocity cell, while the original electron experiences a standard energy loss/deflection as in the case of excitation. The collisional operator for  ionization takes the following form

\begin{equation}
\left(\frac{\delta f_l}{\delta t} \right)_{b}^{ion}  = \left(\frac{\delta f_l}{\delta t} \right)^{ex} (\sigma_b^{ion}) + n_b K^{ion}_b \frac{ \delta(v)}{4 \pi v^2} \delta_{l,0}
\label{eq:en-ion}
\end{equation}
where $\left(\frac{\delta f_l}{\delta t} \right)^{ex} (\sigma_b^{ion})$ is equation \ref{eq:en-ex}, but with $\sigma^{ex}$ replaced with $\sigma^{ion}$, and with the collisional ionization rate coefficient defined (unconventionally) as

\begin{equation}
K^{ion}_b = 4 \pi \int dv v^3 f_0(v) \sigma^{TOT,ion}_b(v).
\end{equation}
Particle sources are then computed using $S_{ion} = \sum_b K^{ion}_b n_b$, and similarly for recombination, while the inelastic collision contribution to $Q$ is similarly calculated by taking the product of each transition rate and the associated transition energy. 

Of course, one operator of the above kinds exists for each possible process of the given kind, i.e. one operator for each excitation process, and one for each ionization process, taking in the appropriate neutral state densities and cross-sections for the given process. The rate coefficients are the same ones used in equation (\ref{eq:CRM}).

Inverse processes (deexcitation and 3-body recombination) are treated using the principle of detailed balance \cite{Bogaerts1998,Colonna2001} to obtain cross-sections. For deexcitation, this gives

\begin{equation}
\sigma_{deex}(i,j,v')=\frac{g_j}{g_i}\frac{v^2}{v'^2}\sigma_{ex}(j,i,v)
\label{eq:detb-deex}
\end{equation}
where $i$ and $j$ are atomic states ($j<i$), and $g_i$ and $g_j$ their statistical weights. For hydrogen these are simply $g_n=2n^2$. Equation (\ref{eq:inel-rel}) defines the velocities, but we use a negative $\epsilon$. 

For 3-body recombination, using the statistical weights of a free electron gas we get for the cross-section

\begin{equation}
\sigma_{3b-recomb}(i,v')\frac{1}{n_e}=\frac{g_i}{2g_1^+}\left(\frac{h^2}{2\pi m_e kT_e}\right)^{3/2}\times\frac{v^2}{v'^2}\sigma_{ion}(i,v),
\label{eq:detb-recomb}
\end{equation}
where $h$ is the Planck constant, $n_e$ and $T_e$ are the electron density and temperature, respectively, and $g_1^+$ is the ion ground state statistical weight (for hydrogen simply $g_1^+=1$).

To calculate the electron fluid mode $R_{en}$ in equation (\ref{eq:el-mom}) we use terms of the form 

$$R_{en}^{ion} =\sum_b\frac{4 \pi}{  3} \int_0^{\infty} \left(\frac{\delta f_1}{\delta t} \right)_{b}^{ion} v^3 dv,$$
where $f_1(v) = - u _e \partial f_0 / \partial v$ (with Maxwellian $f_0$), and similarly for other neutral processes.

\subsubsection{Electron heating operator}

 The implemented diffusive heating operator has the form 

\begin{equation}
 \left(\frac{\partial f_0}{\partial t}\right)_{heating} = \Theta(L_h - x) D(x,t) \frac{1}{3v^2} \frac{\partial}{\partial v}v^2\frac{\partial f_0}{\partial v},
\end{equation}
where $\Theta(L_h-x)$ is the step function designating the heating region. It is easy to check that this operator conserves particle number if $\partial f/\partial v = 0$ on the system boundaries. If we assume a spatially uniform heating, it is easy to show that 

\begin{equation}
D(t) = \frac{W_h(t)}{m_e  \int_0^{L_h} n_e(x,t)dx},
\end{equation}
where $W_h(t)$ is the heat flux entering the SOL over length $L_h$. This is related to the fluid model heating $Q_{ext}$ via $Q_{ext}=W_h/L_h$.

\subsubsection{Particle source operator}

 In order to treat upstream density perturbations, the following electron and ion particle sources (in kinetic mode only) are implemented

\begin{equation}
 \left(\frac{\partial f_0}{\partial t}\right)_{source} = \Theta(L_s - x) F_R(x,t) \left(\frac{m_e}{2\pi k T_{source}}\right)^{3/2}e^{\frac{-mv^2}{2kT_{source}}},
\end{equation}
where $F_R$ is the source rate coefficient

\begin{equation}
 F_R = \frac{\Gamma_{in}}{L_s},
\end{equation}
with $\Gamma_{in}$ being the effective upstream flux. The particles are injected over a length of $L_s$ and with temperature $T_{source}$, which can be the background temperature.

The ion particle source is simply

\begin{equation}
\left(\frac{\partial n_i}{\partial t}\right)_{source} = F_R.
\end{equation}

\subsubsection{Divertor target boundary condition with Legendre polynomials}

 The boundary condition at the divertor target is calculated using the standard logical boundary condition \cite{Procassini1990}, setting the ion and electron fluxes to be equal at the sheath entrance. The logical boundary condition assumes that all electrons with a parallel velocity above some $v_c$ moving towards the target are lost, while all others are reflected. This translates to having a sharp cut-off in the electron distribution function. The challenge when formulating this condition in a Legendre polynomial formalism is the extreme anisotropy that results from it, which would require a high number of harmonics to resolve to a satisfactory level. Fortunately, this number is usually not prohibitively high (see 4.4.). The harmonic content of the ''cut-off'' distribution $f_{cl}$  can be written as a linear combination of known harmonics

\begin{equation}
f_{cl}(v) = \sum_{l'}P_{ll'} f_{l'}(v).
\label{eq:f-div}
\end{equation}
For details on the calculation of the transformation matrix $P_{ll'}$ see Appendix A. Knowing the form of the distribution function, one can solve the ambipolarity condition

\begin{equation}
\frac{4 \pi}{3} \int_0^{\infty} v^3 f_{c1} dv = n_{i,sh} u_{i,sh},
\label{eq:flux-cond}
\end{equation}
where $ n_{i,sh}$ is the extrapolated density at the sheath boundary (see 3.7. below), and $u_{i,sh}$ is the ion velocity at the boundary, given by the Bohm condition

\begin{equation}
u_i \geq c_s = \sqrt{\frac{k(T_e+T_i)}{m_i}},
\end{equation}
where $T_e$ is the electron temperature in the last simulation cell, and $T_i$ is the ion temperature. Solving the ambipolarity condition gives the value of $v_c$, and with it the value of the sheath potential drop $\Delta \Phi=m_e v_c^2 / (2 e)$. The electron distribution harmonics with the correct cut-off velocity can then be used as the dynamically updated boundary condition.

\section{Numerical Methods}

In this section we present numerical details of the SOL-KiT algorithm, starting with the definitions of the normalization scheme and the grids used. An overview of discretization schemes for the various operators in the code follows.

\subsection{Normalization}

The temperature is normalized to some reference value (in eV), while the reference density is assumed to be given in $m^{-3}$. These normalization constants will be refered to as $T_0$ and $n_0 $, respectively. The velocity is normalized to the electron thermal speed $v_{th} =({2 T_0[J]}/{m_e})^{1/2}$, time is normalized to the $90^\circ$ electron-ion collision time $t_0 = {v_{th}^3}/{(\Gamma_{ei}^0n_0\ln\Lambda_{ei}(T_0,n_0)/Z)}$, and the length to the thermal electron-ion collision mean free path $x_0 = v_{th}t_0$. Here $T_0[J]$ denotes the normalization temperature converted from eV to Joules, with

\begin{equation*}
\Gamma_{ei}^0 = Z^2 \Gamma_{ee}^0 = Z^2\frac{e^4}{4\pi(m_e\epsilon_0)^2},
\end{equation*}
and where $\ln\Lambda_{ei}(T_0,n_0)$ is the Coulomb logarithm for electron-ion collisions calculated for the normalization temperature and density (taken from \cite{NRLPF}). All normalized quantities are 

\begin{align*}
\tilde{v} &= \frac{v}{v_{th}}, \quad \tilde{t} = \frac{t}{t_0}, \quad \tilde{x} = \frac{x}{x_0}, \\
\tilde{f_l} &= \frac{f_l}{n_0v_{th}^{-3}}, \quad \tilde{E} = \frac{Eet_0}{m_e v_{th}}, \quad \tilde{q} = \frac{q}{m_e n_0 v_{th}^3}, \\
\tilde{T}_{e,i,g} &= \frac{T_{e,i,g}}{T_0}, \quad \tilde{n}_{e,i,b} = \frac{n_{e,i,b}}{n_0}, \quad  \tilde{u}_{e,i} = \frac{u_{e,i}}{v_{th}}, \\
\tilde{\epsilon} &= \frac{\epsilon}{T_0}, \quad \tilde{\sigma} = \frac{\sigma}{\sigma_0},
\end{align*}
where $\sigma_0 = a_0^2 \pi$ (where $ a_0$ is the Bohr radius) . In the following sections the normalized quantities will be written without the tilde in order to lighten the notation.

\subsection{Grids}

The velocity grid is a uniform or geometric grid of cells with (starting) width $\Delta v_1$ and width multiplier $c_v$, with $N_v$ cell centres distributed as

\begin{align*}
v_1 &= \frac{\Delta v_1}{2}, \quad \Delta v_n = c_v \Delta v_{n-1}, \quad v_n = v_{n-1} + \frac{1}{2}\left(\Delta v_n + \Delta v_{n-1}\right) ,
\end{align*}
while the spatial grid is staggered, i.e. consists of cell centres and boundaries. $N_c$ is the number of cells (cell centres), while $N_x$ is used to denote the total number of spatial points (cells and boundaries), which depends on the boundary conditions of the grid (while $N_c$ is an input parameter).

 In the following text, spatial points with an odd index ($x_1$, $x_3$, etc.) will denote cell centres and those with an even index will denote cell boundaries. The values of these points are determined by

\begin{align*}
x_1 = 0, \quad x_k = x_{k-1} + \frac{\Delta x_{m}^c}{2},
\end{align*}
where $m=k$ if $k$ is odd, or $m=k-1$ if $k$ is even. $\Delta x_{m}^c$ denotes the cell width of the cell whose centre is at $m$. For a uniform grid, this is constant, while for a ``logarithmic'' grid it is an exponential function that starts at a prescribed width for the first cell $dx$, and drops to a prescribed width of the last cell $\Delta x_L$ (with a fixed number of cells $N_c$). 

Variables are positioned on the staggered grid in the following manner:

\begin{itemize}
\item In cell centres:
\begin{itemize}
\item $f_l$ for even $l$
\item Number densities: $n_e$, $n_i$, and $n_b$
\item Electron fluid temperatures $T_e$
\end{itemize}
\item On cell boundaries:
\begin{itemize}
\item $f_l$ for odd $l$
\item $E$-field
\item Ion and electron fluid velocities, $u_i$ and $u_e$
\end{itemize}
\end{itemize}

Variables not evolved in cell centres are linearly interpolated from neighbouring cell boundaries, and vice versa.

\subsection{Timesteps, nonlinear iteration, and vectorization of variables}

 SOL-KiT uses either uniform timesteps of prescribed length $\Delta t$, or rescales $\Delta t$ with $\min(T_e(x)^{3/2}/n_{TOT}(x))$ (where $n_{TOT}$ is the total density of heavy particles). This is a conservative estimate for the shortest Coulomb collision time. In the following text the timestep will be assumed uniform, but generalization to adaptive timesteps is straightforward. Timestepping is done using a first order implicit backwards Euler method. Let $F$ be the vector containing all the evolved quantities, and $M(F)$ the evolution matrix. Then

\begin{equation}
\frac{F^{i+1}-F^i}{\Delta t}=M(F^{i^*})F^{i+1},
\label{eq:implicit1}
\end{equation}
or

\begin{equation}
F^{i+1}=(I - \Delta tM(F^{i^*}))^{-1}F^i,
\label{eq:implicit2}
\end{equation}

Within a given timestep, we use fixed point iteration to solve the non-linear system (\ref{eq:implicit2}), with $M(F^{i^*})$ being evaluated with the solution at the previous iteration.  For the first iteration,$ F^{i^*} = F^i$.  (\ref{eq:implicit2}) is iterated until convergence is established is established within a prescribed tolerance. Implicit variables (those that appear at time $i+1$ on RHS of equation (\ref{eq:implicit1})) in the most important nonlinear terms are given in Tables 1-3.

\begin{table}[htbp]
\centering
\begin{threeparttable}
\caption{Fluid continuity and velocity equation implicit variables in nonlinear terms (including fluid contribution to Amp\`{e}re-Maxwell law)}
\label{tab1}
\begin{tabular}{@{}cccccccccc@{}}
\toprule
Term              & $\frac{\partial (nu)}{\partial x}$ & $S_{ion}\tnote{a}$ & $S_{rec}\tnote{a}$ & $-u\frac{\partial F}{\partial x}$ & $\frac{\partial (nkT)}{\partial x}$ & $-\frac{u}{n}S$ & $R_u$ & $R_{en}$   & $\frac{\partial E}{\partial t} $   \\ \midrule
Implicit variable & $u$                                & $n_b$     & $n_e$\tnote{b}     & $F$                      & $n$                                  & $u$              & $u$     & $n_b$ or $n_e$ & $u$\\ \bottomrule
\end{tabular}%
\begin{tablenotes}
\item [a] Collisional-radiative model uses the same implicit variable as in sources.
\item [b] If using kinetic model this is replaced by $4 \pi \int v^2 f_0 dv$.
\end{tablenotes}
\end{threeparttable}
\end{table}

\begin{table}[htbp]
\centering
\begin{threeparttable}
\caption{Fluid temperature equation implicit variables in nonlinear terms; $ S$ in fifth term refers to same variable as in $S_{ion}$ and $S_{rec}$ of Table \ref{tab1}}

\begin{tabular}{@{}ccccccccc@{}}
\toprule
Term              & $-u\frac{\partial T}{\partial x}$ & $-T\frac{\partial u}{\partial x}$ & $q_T$ & $q_u$ & $-\frac{S}{n}\left[\frac{3}{2}kT - u^2\right]$ & $\frac{u}{n}R_T$ & $\frac{u}{n}R_u$ & $\frac{u}{n}R_{en}$ \\ \midrule
Implicit variable & $T$                               & $u$                              & $T$   & $u$   & $S$                                           & $T$              & $u$              & $n_b$ or $n_e$      \\ \bottomrule
\end{tabular}%
\end{threeparttable}

\end{table}

\begin{table}[htbp!]
\centering
\begin{threeparttable}
\caption{Electron kinetic equation implicit variables in nonlinear terms; Coulomb collision terms for $f_0$ written out in more detail to avoid confusion (see below)}

\begin{tabular}{@{}cccccc@{}}
\toprule
Term              & $E\frac{\partial f}{\partial v}$ & $\left(\frac{\delta f_0}{\delta t}\right)_{e-e}$          & $\left(\frac{\delta f_{l >0}}{\delta t}\right)_{e-i}^{stationary}$ & $\left(\frac{\delta f_{l>0}}{\delta t}\right)_{e-i}^{moving}$ & $\left(\frac{\delta f_{l}}{\delta t}\right)_{e-n}$ \\ \midrule
Implicit variable & $E$                              & $C^{i^*},D^{i^*},f_0^{i+1},\partial f_0^{i+1}/\partial v$ & $f_l$                                                              & $f_l$, and $n_i$ in $I(F_l)$,$J(F_l)$                              & $f_l$                                              \\ \bottomrule
\end{tabular}%
\end{threeparttable}
\end{table}

The structure of the variable vector (with all variables present) is the following: 

\begin{equation}
F =  \begin{pmatrix}F_{loc}(x_1)\\F_{loc}(x_2)\\ \vdots \\ F_{loc}(x_{N_x}) \end{pmatrix} ,
\label{eq:F-vec-sp}
\end{equation} 
where $F_{loc}(x_k)$ is the (spatially) local subvector for quantites at point $x_k$ given as

\begin{equation}
F_{loc}(x_k) =  \begin{pmatrix}f_{l_{max}}(x_k)\\f_{l_{max - 1}}(x_k)\\ \vdots \\ f_{0} (x_k)
\\ n_1 (x_k)\\ n_2 (x_k) \\ \vdots \\ n_{N_n} (x_k)\\ E (x_k)\\ n_e(x_k) \\ u_e (x_k) \\ T_e(x_k) \\ n_i(x_k) \\ u_i (x_k)\end{pmatrix} ,
\label{eq:F-vec-loc}
\end{equation}
where $N_n$ is the total number of neutral states tracked, $l_{max}$ is the highest resolved harmonic, and $f_l(x_k)$ is the $l$-th harmonics subvector

 \begin{equation}
f_{l}(x_k) =  \begin{pmatrix}f_l(x_k,v_1) \\ f_l(x_k,v_2)\\ \vdots \\ f_l(x_k,v_{N_v})\end{pmatrix}.
\label{eq:F-vec-l}
\end{equation}

Note that when running in kinetic mode, the vector does not contain electron fluid quantities, while when the code is running with fluid electrons the distribution function harmonics are not evolved, but are updated after each timestep to be a slowly drifting Maxwellian with current temperature, density, and electron fluid velocity.

From knowing the input parameters $N_c$ (and its derived parameters $2N_c - 1\leq N_x \leq 2(N_c+2) - 1$), $l_{max}$, $N_n$, and $N_v$ we can calculate the total vector length (for a kinetic run) as

\begin{equation}
N_{total} = N_x\left((l_{max} + 1)N_v + N_n + 3\right).
\end{equation}
For a representative system  with $l_{max} = 5$, $N_v= 80$, $N_n = 30$, and $N_x = 128$ this gives $N_{total} = 55424$.

To solve the above matrix system, we use the MPI and PETSc libraries \cite{petsc-web-page,petsc-user-ref,petsc-efficient}. Domain decomposition is done in the spatial dimension with as close to even distibution of grid points between processors as possible.  

\subsection{Velocity and spatial derivative discretization}

 The velocity space derivatives appearing in the various kinetic operators are all implemented using a central difference scheme:

\begin{align*}
\frac{\partial^2 F}{\partial v^2} (x_k,v_n)&= \frac{1}{\Delta v_n}\left[ \frac{F(x_k,v_{n+1}) - F(x_k,v_{n})}{v_{n+1}-v_n}-\frac{F(x_k,v_{n}) -F(x_k,v_{n-1})}{v_n-v_{n-1}}\right], \\
\frac{\partial F}{\partial v} (x_k,v_n)  &= \frac{F(x_k,v_{n+1}) - F(x_k,v_{n-1})}{v_{n+1}-v_{n-1}}, \\
\end{align*}
where $F$ is any velocity space function. 

Spatial derivatives are mostly discretized using an analogous  central difference scheme to the above velocity space one. The only exceptions are the advection terms in the momentum and temperature equations (eq. (\ref{eq:el-mom}),(\ref{eq:el-temp}),(\ref{eq:ion-mom})), where an upwind scheme can be used instead of the central difference. The upwind scheme used is simply

\begin{equation}
\frac{\partial F}{\partial t} (x_k) = - u (x_k) \frac{\partial F}{\partial x} (x_k),
\end{equation}
where if $u(x_k) \ge 0$

\begin{equation}
\frac{\partial F}{\partial x} (x_k) = \frac{F(x_k) - F(x_{k-1})}{x_k - x_{k-1}},
\end{equation}
and if $u(x_k) < 0$

\begin{equation}
\frac{\partial F}{\partial x} (x_k) = \frac{F(x_{k+1}) - F(x_{k})}{x_{k+1} - x_{k}},
\end{equation}
where if $F$ is not evolved on $x_{k \pm 1}$ we use use $x_{k \pm 2}$ instead (this is the case for temperature advection in (\ref{eq:el-temp})).

Finally, the diffusion type derivatives are given by

\begin{align*}
\frac{\partial}{\partial v}\left(A\frac{\partial F}{\partial v}\right) (x_k,v_n)&= \frac{1}{\Delta v_n}[A(x_k,v_{n+1/2}) \frac{F(x_k,v_{n+1}) - F(x_k,v_{n})}{v_{n+1}-v_n}\\ &- A(x_k,v_{n-1/2})\frac{F(x_k,v_{n}) -F(x_k,v_{n-1})}{v_n-v_{n-1}}], \\
\frac{\partial}{\partial x}\left(A\frac{\partial F}{\partial x}\right) (x_k)&= \frac{1}{x_{k+2} - x_{k-2}}[A(x_{k+1}) \frac{F(x_{k+2}) - F(x_{k})}{x_{k+2}-x_k}\\ &- A(x_{k-1})\frac{F(x_k) -F(x_{k-2})}{x_k-x_{k-2}}],
\end{align*}
where the velocity grid boundaries $v_{n+1/2}$ are given as $v_{n+1/2} = \sum_{l=1}^n \Delta v_l$. 

The only simple derivatives that do not obey the above are in the velocity advection terms, where in equations (\ref{eq:G-l}) and (\ref{eq:H-l}) we use the conservative form 

$$\frac{\partial F}{\partial v} (x_k,v_n)  = \frac{F(x_k,v_{n+1/2}) - F(x_k,v_{n-1/2})}{\Delta v_n},$$
where $F(x_k,v_{n+1/2})$ is obtained through linear interpolation.

The following velocity space boundary conditions are assumed for the distribution function harmonics

\begin{equation}
f_0(x_k,0) = \frac{f_0(x_k,v_1) - f_0(x_k,v_2)v_1^2/v_2^2}{1-v_1^2/v_2^2},
\end{equation}

\begin{equation}
f_{l>0}(x_k,0) = 0,
\end{equation}
i.e. $f_0$ at $v=0$ is quadratically extrapolated (this is used whenever $f_0$ at $v=0$ is required) and higher harmonics are set to $0$. At the velocity space boundary after $v_{N_v}$ all $f_l$ and their derivatives are assumed to be zero.

\subsection{Coulomb collision operators}

 Here, Coulomb collision operator discretization will briefly be presented to supplement the material in previous sections. The method used for the electron-electron collisions for $l=0$ is the Chang-Cooper-Langdon\cite{Chang1970,Epperlein1994,Langdon1981} method. The implementation in SOL-KiT is very similar to that in IMPACT\cite{Kingham2004}, and as such we will leave out some of the details. We write the collision operator as a divergence of fluxes $F$

\begin{equation}
C_{ee0}^{i+1}(x_k,v_n) = \frac{A_{ee}^0 \ln\Lambda_{ee}(T_e^{i}(x_k),n_e^{i}(x_k))}{v_n^2}\frac{F^{i+1}(x_k,v_{n+1/2}) - F^{i+1}(x_k,v_{n-1/2})}{\Delta v_n},
\end{equation} 
where $A_{ee}^0 = \Gamma_{ee}^0 n_0 t_0/ v_t^3=1/(Z \ln\Lambda_{ei}(T_0,n_0))$, and the flux is given by 

\begin{align}
F^{i+1}(x_k,v_{n+1/2}) & \notag = C^{i^*}(x_k,v_{n+1/2})f_0^{i+1}(x_k,v_{n+1/2})\\ &+ D^{i^*}(x_k,v_{n+1/2})\frac{ f_0^{i+1}(x_k,v_{n+1}) - f_0^{i+1}(x_k,v_{n}) }{v_{n+1}-v_n}.
\end{align} 
$f_0^{i+1}(x_k,v_{n+1/2})$ is then calculated using a special weighted interpolation, which ensures relaxation to a Maxwellian

\begin{align}
f_0^{i+1}(x_k,v_{n+1/2}) & \notag= (1-\delta^{i^*}(x_k,v_{n+1/2}))f_0^{i+1}(x_k,v_{n+1}) \\ &+ \delta^{i^*}(x_k,v_{n+1/2})f_0^{i+1}(x_k,v_n), \\
\delta^{i^*}(x_k,v_{n+1/2}) &= \frac{1}{W^{i^*}(x_k,v_{n+1/2})} - \frac{1}{\exp[W^{i^*}(x_k,v_{n+1/2})] - 1}, \\
W^{i^*}(x_k,v_{n+1/2}) &= (v_{n+1}-v_n)\frac{C^{i^*}(x_k,v_{n+1/2})}{D^{i^*}(x_k,v_{n+1/2})}.
\end{align} 

The friction coefficient is

\begin{equation}
C^{i^*}(x_k,v_{n+1/2}) = 4 \pi \sum_{l=1}^n f_0^{i^*}(x_k,v_l)v_l^2\Delta v_l.
\end{equation} 
As previously noted in the Section 2.1., we chose to write the diffusion coefficient in a way different from what is usually in the literature. This allows discretization that conserves energy as well

\begin{equation}
D^{i^*}(x_k,v_{n+1/2}) = \frac{4 \pi}{v^*_{n+1/2}} \sum_{l=1}^n v_l^2\left[\sum_{m=l}^{N_v-1}f_0^{i^*}(x_k,v_{m+1/2})v_{m+1/2}(v_{m+1}-v_m)\right]\Delta v_l,
\end{equation} 
where  $v^*_{n+1/2} = (v_n + v_{n+1})/2$. Boundary conditions for the $C$ and $D$ coefficients are taken in such a way to conserve particle density (see \cite{Kingham2004}). The resulting submatrix from this operator is tridiagonal. Thus we have an iterative method which conserves particles and energy (up to nonlinear iteration tolerance), and which relaxes the distribution to a Maxwellian in the absence of other driving forces.

The electron-electron collision terms for higher $l$ (equation (\ref{FP-l>0-ee}) have the same normalization constant as that for $l=0$. Here we use a discretization method similar to that in OSHUN \cite{Tzoufras2011}, and will hence again, for the sake of brevity go over only the most important elements.
The first three terms in  (\ref{FP-l>0-ee}) produce a tridiagonal matrix when discretized, while the remaining terms produce an upper and lower triangular submatrix due to the the $I$ and $J$ integrals, which are discretized as 

\begin{equation}
I_j[f_l](x_k,v_n) = 4 \pi \begin{pmatrix} \left(\frac{v_1}{v_n}\right)^j v_1^2 \Delta v_1 \\ 
\left(\frac{v_2}{v_n}\right)^j v_1^2 \Delta v_2 \\ \vdots \\
\left(\frac{v_{n-1}}{v_{n}}\right)^j v_{n-1}^2 \Delta v_{n-1} \\
\frac{1}{2}v_n^2 \Delta v_n \\ 0 \\ \vdots \\ 0\end{pmatrix}^T \cdot 
 \begin{pmatrix} f_l(x_k,v_1) \\ \vdots \\ f_l(x_k,v_N)\end{pmatrix},
 \label{eq:Ij-fd}
\end{equation} 

\begin{equation}
J_j[f_l](x_k,v_n) = 4 \pi \begin{pmatrix}  0 \\ \vdots \\ 0 \\
\frac{1}{2}v_n^2 \Delta v_n \\ 
\left(\frac{v_{n+1}}{v_{n}}\right)^j v_{n+1}^2 \Delta v_{n+1} \\ \vdots \\
\left(\frac{v_{N_v-1}}{v_{n}}\right)^j v_{N_v-1}^2 \Delta v_{N_v-1} \\
\left(\frac{v_{N_v}}{v_{n}}\right)^j v_{N_v}^2 \Delta v_{N_v} \end{pmatrix}^T \cdot 
 \begin{pmatrix} f_l(x_k,v_1) \\ \vdots \\ f_l(x_k,v_N)\end{pmatrix}.
 \label{eq:Jj-fd}
\end{equation}

For $l=1$ the discretized e-e collision operator does not numerically conserve momentum, unfortunately, as the discertized form loses the partial integration properties that would analytically conserve momentum. However, the numerically lost momentum is transfered to the ions, and total momentum is thus conserved. 

The electron-ion operator for stationary cold ions is trivial, and is discretized straightforwardly, while the moving ion operator is discretized similarly to the e-e operator for higher $l$, with the exception of having mainly a tridiagonal component (with terms containing $F_0$), and a part (terms containing $F_l$) where ion density is the implicit variable (if working on a cell boundary, it is implicitly interpolated) .  For the second part we need the previously presented $I(F_l)$ and $J(F_l)$ integrals for cold ions ((\ref{Ij-ci}) and (\ref{Jj-ci})). For a given lagged ion velocity $u_i^{i^*}(x_k)$ and density $n_i^{i+1}(x_k)$ these integrals in discrete form are

\begin{equation}
I_j^{i+1}[F_l](x_k,v_n)= (2l+1) n_i^{i+1}(x_k) \frac{u_i^{i^*}(x_k)^j}{v_n^j} \Theta(v_k-u_i^{i^*}(x_k)),
\label{Ij-ci-fd}
\end{equation} 

\begin{equation}
J_j^{i+1}[F_l](x_k,v_n)= (2l+1) n_i^{i+1}(x_k) \frac{u_i^{i^*}(x_k)^j}{v_n^j} \Theta(u_i^{i^*}(x_k) - v_k).
\label{Jj-ci-fd}
\end{equation}

\subsection{Electron-neutral collision numerics}

 In this section we briefly go over the basic properties of the elastic electron-neutral collision operator before moving on to the important conservative discretization of inelastic electron-neutral collisions.

\paragraph{Elastic collisions} The discretization of elastic collisions borrows greatly from the Coulomb collision operators. Equation (\ref{eq:en-el-0}) is discretized in a way similar to the Chang-Cooper-Langdon scheme. However, since the integral is linear in $f_0$, there is no need for special interpolation ($f_0$ and $\sigma$ are simply linearly interpolated), and just the flux formalism was used, with the $C$ and $D$ coefficients (here left unnormalized) being

\begin{equation}
C(v_{n+1/2}) =n_b  v_{n+1/2}^4\sigma^{el}(v_{n+1/2}),
\end{equation} 

\begin{equation}
D(v_{n+1/2}) =n_b v_{n+1/2}^3\sigma^{el}(v_{n+1/2})\frac{k T_g}{m_e},
\end{equation} 
where

\begin{equation}
\sigma^{el}(v_n) = \int d\Omega (1 - \cos\chi) \sigma^{el}(\chi,v_n).
\end{equation}
Since no differential cross section data is available, this is simply set to the constant cross-section (\ref{eq:sigma_el}).

The above discretization produces an operator that conserves particle number and relaxes the electron $f_0$ to a Maxwellian with temperature $T_g$ (within finite grid effects). For higher $l$, equation (\ref{eq:en-el-l}) is discretized in a straightforward way.

\subsubsection{Conservative discretization scheme for inelastic collisions}

 In order to streamline the arguments in the next sections, let us introduce the following notation. If we label inelastic processes with transition energy $\epsilon$ as $\Pi$, we can label the corresponding superelastic (inverse) processes as $\Pi^{-1}$ (with transition energy $-\epsilon$). Electron energy and particle conservation are governed by the $l=0$ harmonic, which we will denote simply as $f$ in the following derivation. Then the value of $f$ in the $n$-th velocity cell centre can be written as $f_n$, while the total cross-section notation for any given process will be labeled as $\sigma_n$.

Equation (\ref{eq:en-ex}) contains two terms, a loss/emission term and a gain/absorption term. While the first term is defined on the used velocity grid, the second term requires evaluation of the distribution function on (most likely) non-existant points $\alpha(v)  v$ (where $\alpha = (1\pm 2\epsilon/mv^2)^{1/2}$). A straightforward interpolation here fails to produce a discretization that conserves particles and energy. In order to develop such a scheme we start by writing the collision operator in the emission/absorption form 

\begin{equation}
C_n = - E_n + A_n,
\end{equation}
where

\begin{equation}
E_n =  n_{neut} v_n  f_n \sigma_n,
\label{eq:E-n}
\end{equation}
and 

\begin{equation}
A_n =  \sum_m W_{nm} E_m,
\label{eq:A-n}
\end{equation}
where $W_{nm}$ are weights determining the contribution of the emission from cell $m$ to the absorption in cell $n$. For particle conservation, we want the number of particles emitted by a single cell $m$ - $4 \pi E_m v_m^2\Delta v_m$ to be absorbed exactly by some other cells $n$ - $4 \pi W_{nm} E_mv_n^2\Delta v_n$. After tidying up, the resulting particle conservation condition is

\begin{equation}
 v_m ^ 2 \Delta v_m = \sum_ n W_{nm} v_n ^2 \Delta v_n.
 \label{eq:part_cons}
\end{equation}

For energy conservation, we note that all energy being emitted via collisions from one cell needs to either be lost to the energy of internal atomic states or be absorbed by some cells $n$. In a similar way to the above, one can cancel distribution functions and cross-sections to obtain the energy conservation condition

\begin{equation}
v_m ^ 2  \Delta v_m \left( v_m ^2 \mp \epsilon  \right) = \sum_nW_{nm}  v_n^4 \Delta v_n. 
\end{equation}

Finally, it is useful to write down the numerical version of detailed balance for the collisional cross-section of the inverse process. Here we show the result for de-excitation ($\Pi = ex$, $\Pi^{-1} = deex$), with the same procedure applicable to recombination. Detailed balance implies that for a Maxwellian distribution of electrons and for a Boltzmann distribution of excited states the rates of excitation and de-excitation become equal. This can be written as

 \begin{equation}
n_l \sum_m f_m \sigma_m^{ex} v_m ^3 \Delta v_m = n_u \sum_n f_n \sigma_n^{deex} v_n^3 \Delta v_n,
\end{equation}
where $n_l$ and $n_u$ denote the densities of the lower and upper excited states, respectively. Using a Maxwellian for $f$, relating $n_u$ and $n_l$ via the Boltzmann distribution, and utilizing equation (\ref{eq:part_cons}) we obtain

\begin{equation}
 \sigma_n^{deex} = \frac{g_l}{g_u} e^{\epsilon/T}  \sum_m W_{nm} \sigma_m^{ex} \frac{v_m}{v_n}  e^{-\left(v_m^2 - v_n^2\right)/T}. 
 \label{eq:det-balance}
\end{equation}
Note that this depends on the temperature and the excitation weights, contrary to the analytical version from equation (\ref{eq:detb-deex}). This is due to the discrete nature of the grid, where energy differences between absorbing and emitting cells do not have to be equal to the analytical value $\epsilon$. 

\subsubsection{Two-absorber mapping}

 Note that in the above conservation condition we haven't specified the summation range for $n$. The simplest way to do this is the following. For each emitter $m$ we choose exactly two (consecutive) absorber cells, $n_1$ and $n_2$ such that $\sqrt{v_m^2 \mp \epsilon}$ lies between  $v_{n_1}$ and $v_{n_2}$. We will refer to these two points as the ideal absorber pair. This way we do not need any further instructions on partitioning the emitted particles and energy, and can proceed to calculate weights. Since for any given $m$ there are only two absorbing cells, we can denote the weights simply as $W_1^m$ and $W_2^m$, and solving the conservation conditions we get

\begin{equation}
 W_1^m  = \frac{v_m ^ 2 \Delta v_m}{v_{n_1(m)} ^2 \Delta v_{n_1(m)}}\left( 1 - \frac{v_m^2 - v_{n_1(m)}^2 \mp \epsilon}{ v_{n_2(m)}^2- v_{n_1(m)}^2}\right) , 
 \label{eq:w1-2a}
\end{equation}

\begin{equation}
 W_2^m  = \frac{v_m ^ 2 \Delta v_m}{v_{n_2(m)} ^2 \Delta v_{n_2(m)}}\frac{v_m^2 - v_{n_1(m)}^2 \mp \epsilon}{ v_{n_2(m)}^2- v_{n_1(m)}^2} . 
 \label{eq:w2-2a}
\end{equation}

\begin{figure*}
\centering
\includegraphics[width=0.75\textwidth]{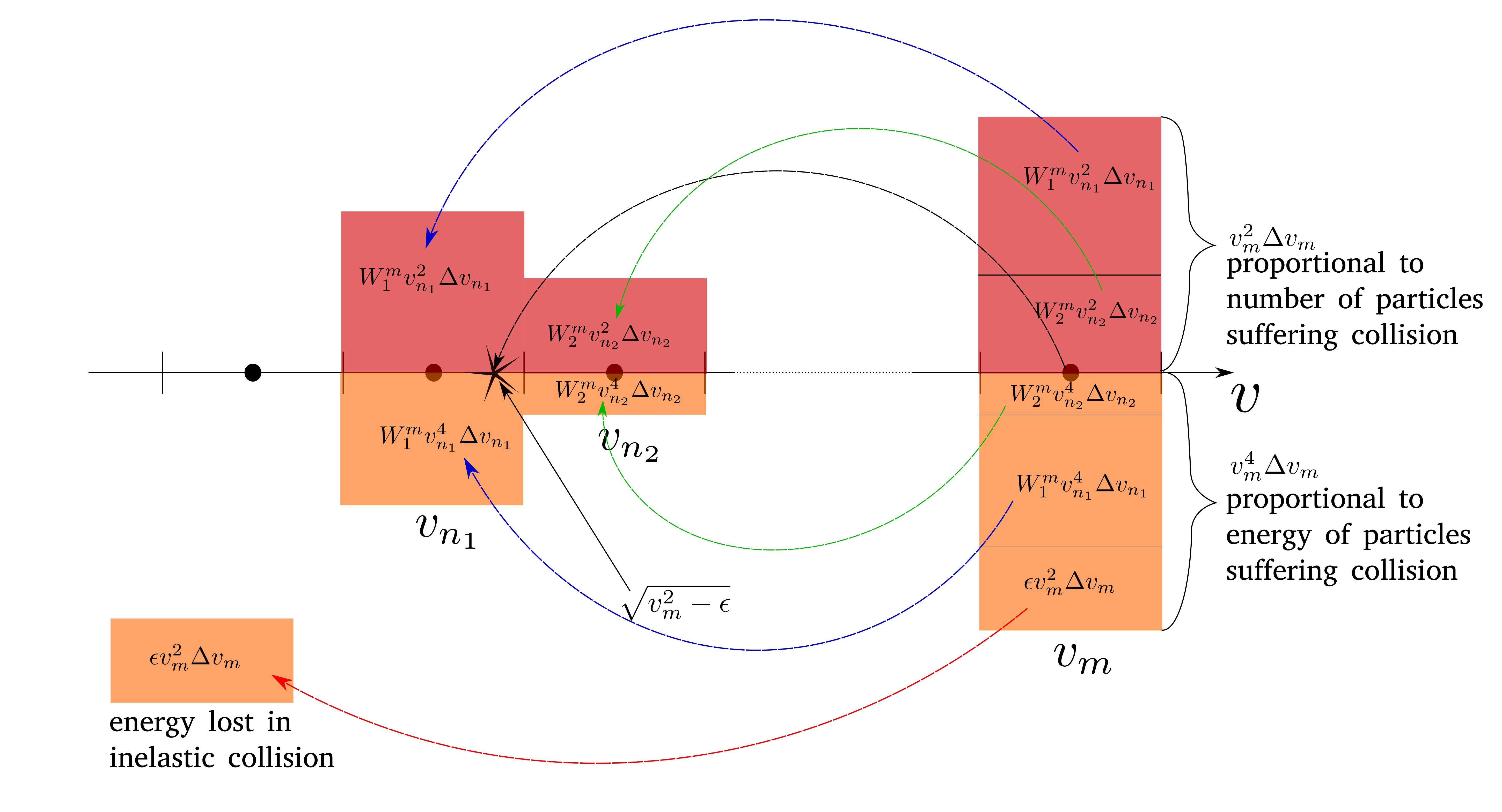}
\caption{Two-absorber mapping for an excitation collision; particles and energy emitted by higher energy cell distributed among the two absorbers, while a portion of energy is lost to internal energy states of collision target (atom) - red line. Blue and green lines denote emission to first and second cell of absorber pair, respectively. The black star shows the location of the analytical absorption point.}
\label{fig:2abs}
\end{figure*}

Figure \ref{fig:2abs} shows an illustration of this mapping for an excitation collision with both particle and energy transfer obeying conservation conditions.

To establish that this scheme can reduce to the analytical result, we note two properties we expect in the analytic limit. Firstly $n_1 \rightarrow n_2 = n$, i.e. each emitting point has one and only one absorbing point. This can be done by letting one of the weights tend to $0$. For the sake of this illustrative argument, let that be $W_1^m$, from which we see that  the second fraction in (\ref{eq:w2-2a}) tends to unity. Then, we note that the first fraction will tend to $\alpha(v_{n})^2 \Delta v_m / \Delta v_{n}$ (since $v_m = \alpha(v_n) v_n$ in analytical limit). Taking the differential of (\ref{eq:inel-rel}) points us toward a grid that satisfies 

$$\Delta v_m / \Delta v_{n} \rightarrow 1/\alpha(v_{n})$$
which finally yields the limits

\begin{equation*}
 W_1^m  \rightarrow 0, \quad W_2^m \rightarrow \alpha(v_{n}).
\end{equation*}
Noting that we have chosen to write the gain term as a sum of weighted loss terms, another $ \alpha$ factor can be extracted from the single emitter cell velocity in the absorption term, and the analytical $\alpha^2$ form from (\ref{eq:en-ex}) is recovered. 

At first glance, this mapping looks good, we have found pairs of absorbers and weighted their absorption terms to conserve particles and energy, but after a closer look at eq. (\ref{eq:det-balance}) a potential problem reveals itself. Suppose that for some $n$ and every $m$ the weights $W_{nm} =0$ in an excitation process. In this case $\sigma_n^{deex}=0$ for the corresponding deexcitation process. In other words, if cell $n$ does not absorb in process $\Pi$, it will not emit in $\Pi^{-1}$, and gaps are left on our velocity grids where cells that should emit do not. This is not physical, and the mapping needs to be refined.

\subsubsection{Two-absorber mapping with pair partition}

 In the previous section we have associated every cell (emitter) $m$ with an ideal absorber pair $(n_1(m),n_2(m))$. We define the absorber list $\mathcal{A}^{\Pi}_m$ as a list of absorber pairs of point $m$ in process $\Pi$. The two-absorber mapping implies $\mathcal{A}^{\Pi}_m = \left\{ (n_1(m),n_2(m))\right\}$, where the only pair is the ideal absorber pair. As noted above, this produces gaps in the the velocity grid where cells do not emit in the inverse process $\Pi^{-1}$. 

In order to fill out the aforementioned gaps, we must potentially include absorbers other than those in the ideal pair. This can be done by looking at all cells $n$ that aren't in the ideal absorber pair, but whose $\sqrt{v_n^2 \pm \epsilon}$ (note the change in sign!) falls into cell $m$. We then refer to $m$ as the ideal emitter for cells $n$.

We then move to generate absorber pairs that would include the non-ideal absorbers $n$, while satisfying the original constraint, namely that $\sqrt{v_m^2 \mp \epsilon}$ is between the points of each new pair. In SOL-KiT this is done by always pairing non-ideal absorbers with one of the ideal absorbers. If we denote $P_m \ge 1$ as the number of (both ideal and non-ideal) absorber pairs of cell $m$, the absorber list becomes $\mathcal{A}^{\Pi}_m = \left\{ (n_1^p(m),n_2^p(m)), p=1,...,P_m\right\}$, where we label the first and second cell in pair $p$ as $n_1^p(m)$ and $n_2^p(m)$, respectively.

\begin{figure*}[htbp]
\centering
\includegraphics[width=0.6\textwidth]{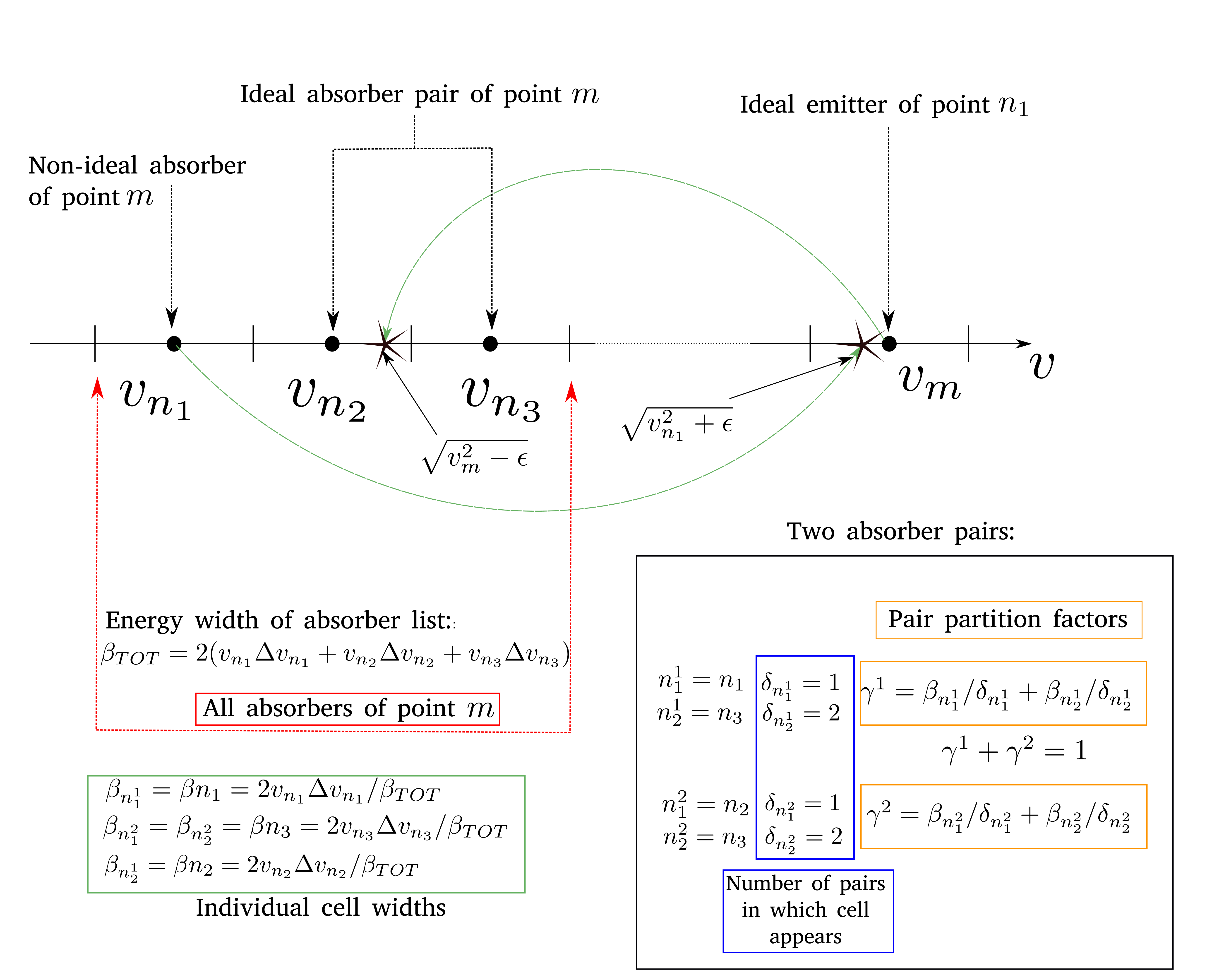}
\caption{Pair partition calculation example for an excitation collision; emitter cell has a list of three absorbers, grouped into two pairs, with emitted energy and particles partitioned according to factors $\gamma^1$ and $\gamma^2$}
\label{fig:2abs-pp}
\end{figure*}

 This allows us to use the previous solution, but we require a way to partition the emitted energy/particles among pairs. We do this by defining a total energy width of the entire absorber list and normalizing the energy widths of each cell within the list to the total energy width of the list:

\begin{equation*}
\beta_{TOT}^m =  \sum_{n \in\mathcal{A}^{\Pi}_m} 2v_n\Delta v_n, \quad \beta_n =  2 v_n \Delta v_n / \beta_{TOT}^m.
\end{equation*}
Then we define $\delta_n$ as the number of pairs in $\mathcal{A}^{\Pi}_m$ which contain point $n$. We can then define

\begin{equation}
\gamma^p = \frac{\beta_{n_1^p}}{\delta_{n_1^p}} +  \frac{\beta_{n_2^p}}{\delta_{n_2^p}}
\end{equation}
to be the fraction of the emission given to pair $p$. It is trivial to check that the sum of all $\gamma^p$ for a given absorber list is equal to unity, as one would expect. An example with an absorber list with three distinct cells is given in Figure \ref{fig:2abs-pp}.

Then we can go through the same process as the one for the two-absorber case, except the final results for $W_1^m$ and $W_2^m$ will now be functions of the pair $p$ for which they have been calculated, and will have an extra $\gamma^p$ factor multiplying the previous simple results (see below). To then calculate the total absorption term for a given cell $n$, it should be summed over each pair the cell belongs to, i.e.

\begin{equation}
A_n =  \sum_m \sum_{p} W_{nm}^p E_m.
\end{equation}
This way we have both ensured particle and energy conservation, as well as a reasonably physical numerical detailed balance condition. The weights in this case are given by

\begin{equation}
 W_{n_1^p(m)m} = \gamma_m^p \frac{v_m ^ 2 \Delta v_m}{v_{n_1^p(m)} ^2 \Delta v_{n_1^p(m)}}\left( 1 - \frac{v_m^2 - v_{n_1^p(m)}^2 \mp \epsilon}{ v_{n_2^p(m)}^2- v_{n_1^p(m)}^2}\right) , 
\end{equation}

\begin{equation}
 W_{n_2^p(m)m}  = \gamma_m^p  \frac{v_m ^ 2 \Delta v_m}{v_{n_2^p(m)} ^2 \Delta v_{n_2^p(m)}}\frac{v_m^2 - v_{n_1^p(m)}^2 \mp \epsilon}{ v_{n_2^p(m)}^2- v_{n_1^p(m)}^2} . 
\end{equation}

Using the above method for every process produces both a transition mapping and weights for each one. These depend solely on the grid and the inelastic processes being considered, and as such do not change during a simulation. The final discretized (unnormalized) form of the inelastic collision integral for particle conserving collisions is then

\begin{align}
\left(\frac{\delta f_l}{\delta t} \right)^{inel,i+1}_{b\rightarrow b'}(x_k,v_n) & \notag= -   v_nn_{b}^{i^*}(x_k) [\sigma^{TOT}_{b\rightarrow b'}(v_n) f_l^{i+1}(x_k,v_n) \\ 
& - \sum_m \sum_p W_{nm}^p (\sigma^{TOT}_{b\rightarrow b'}(v_m) -  \sigma^{(l)}_{b\rightarrow b'}(v_m))f_l^{i+1}(x_k,v_m)].
\end{align}

\subsection{Divertor target boundary condition discretization}

 In order to implement the boundary condition at the divertor target (not explicitly present on the spatial grid), as was implied in the Section 2.1.6., we require knowledge of the forward going electron distribution function. We reconstruct it by extrapolating harmonics from cells leading up to the boundary

\begin{equation}
f_{\text{odd } l}^{\text{forward,target}} = \frac{n_e^2(x_{N_x})}{n_e^2(x_{N_x - 1})}f_{\text{odd } l}^{i^*}(x_{N_x-1}), \quad f_{\text{even } l}^{\text{forward,target}} = \frac{n_e(x_{N_x})}{n_e(x_{N_x - 1})}f_{\text{even } l}^{i^*}(x_{N_x}),
\end{equation}

where we scale the harmonics by the ratio of the electron densities at spatial cells $N_x$ and $N_x-1$. The choice to perform this sort of extrapolation, and not a linear one, comes from the danger of large density gradients (which are often present at the divertor target) to produce negative extrapolated densities when extrapolated linearly.

\begin{figure*}
\centering
\includegraphics[width=0.5\textwidth]{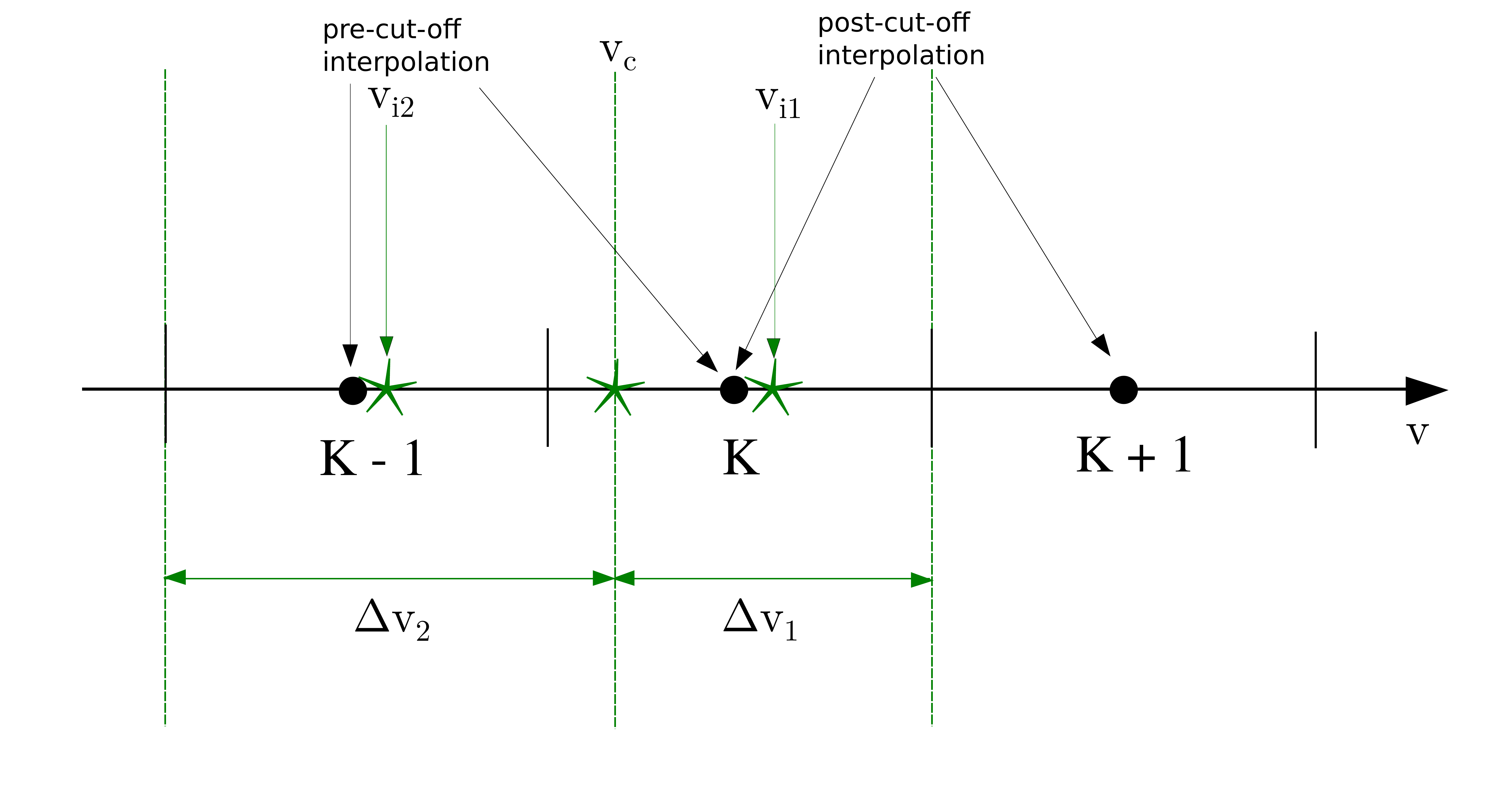}
\caption{The interpolation region on the velocity grid; cell $K$ contains the cut-off, and cells $K$ and $K-1$ are replaced with interpolated cells with centres at $v_{i1}$,$v_{i2}$ and widths $\Delta v_1$,$\Delta v_2$}
\label{fig:disc}
\end{figure*}

Knowing the forward going distribution allows us to calculate the cut-off distribution using equation (\ref{eq:f-div}), while equation (\ref{eq:flux-cond}) imposes $v_c$. However, due to the discretization of velocity space, this will not be one of the resolved cell centres, and we are required to interpolate in order to calculate the precise cut-off. A diagram of the interpolation region is given in Figure  \ref{fig:disc}.

The interpolation replaces cell $K$  containing the cut-off and its preceding cell $K-1$ with two new cells with centres

$$v_{i1} = \frac{v_{K+1/2} + v_c}{2}, \quad v_{i2} = \frac{v_{K-3/2} + v_c }{2},$$
and with widths

$$\Delta v_{1} = v_{K+1/2} - v_c, \quad \Delta v_{2} =v_c - v_{K-3/2}.$$

The electron flux to the boundary (given in equation (\ref{eq:flux-cond})) is then calculated using this updated grid, while linearly interpolating the $f_1$ component of the cut-off distribution in the two new cells. The electron flux can then be matched to the ion flux using a bisection or false position method (the latter being implemented in the current version of SOL-KiT) to find $v_c$ with a desired accuracy.

\section{Benchmarking SOL-KiT operators}

 A number of verification tests have been performed using SOL-KiT, aimed at checking the properties of various implemented operators. The test details are presented in the following sections, while an overview of the tests and their scope is given in Table \ref{tab:all_runs}. In all tests we consider hydrogenic species, ie. $Z=1$.

\begin{table}[]
\caption{List of performed test runs and sets of runs with their target operators. }
\label{tab:all_runs}
\centering
\resizebox{0.6\textwidth}{!}{%
\begin{tabular}{@{}ll@{}}
\toprule
Run/Set                    & Targeted operators                                                        \\ \midrule
Runs 1-4                   & e-i and e-e collisions, Vlasov, Maxwell                                   \\
Runs 5-6                   & fluid ion and electron operators                                          \\
Set 1                      & e-i and e-e collisions, Vlasov, Maxwell, high $l$                         \\
Runs 7-8                   & kinetic e-n inel. collisions, CRM                                         \\
Run 9                      & fluid e-n inel. collisions, CRM, detailed balance                         \\
Run 10                     & e-i and e-e collisions, Vlasov, Maxwell, kinetic e-n inel. collisions, CRM \\
Run 11                     & fluid ion operators - advection                                           \\
Set 2                      & fluid ion operators - charge-exchange friction                                                  \\
Set 3                      & divertor boundary condition - analytic limit                              \\ 
Set 4 &divertor boundary condition, high $l$              \\ \bottomrule
\end{tabular}%
}
\end{table}

\subsection{Heat conduction tests}

In order to test a number of operators we performed both local tests with different scenarios, as well as non-local perturbation tests.

\paragraph{Local heat conduction tests}Two types of tests were used to verify the local limit of heat conduction in SOL-KiT. The first type is a small perturbation test on a periodic grid, with the initial temperature given by

\begin{equation}
T(x) = T_0 + T_1 \sin\left(2\pi\frac{x}{L}\right),
\end{equation}
where $L$ is the total length of the system, given by $L=N_c \Delta x$. The density was set to $n_0 = 10^{19}m^{-3}$, and the rest of the simulation parameters are 

$$ T_0 = 100 \text{eV},\quad T_1 = 0.05 \text{eV},$$
$$ N_c = 64, \quad \Delta x = 150 \ x_0,\quad N_v = 120,$$
with $\Delta v$,$\Delta t$, and the total number of timesteps being able to vary between runs. The total time simulated was always set to $30 \ t_0$, i.e. to 30 electron-ion collision times. In all runs the full set of Coulomb collision operators was active, and the highest harmonic number was kept at $l_{max}=1$. The calculated conduction heat flux was compared to the Spitzer-H\"{a}rm (SH) heat flux $q_{SH}$ (equation (\ref{eq:SH-q})), with the heat conductivity for $Z=1$ in SOL-KiT normalized units being $\kappa= 0.6 \sqrt{\pi}.$

The ratio of the calculated heat flux and the reference SH flux at the end of each run would be averaged along the simulation domain, and these are the results presented below. 
Three pure kinetic runs (with initial $f_1$ set to 0, and no fluid ion operators active) were performed. The reference run (Run 1) with $\Delta t = 0.1 \ t_0$, $\Delta v = 0.1 \ v_{th}$, $N_t = 300$ produces an average heat flux of $q =(0.988296 \pm 1\times 10^{-6})q_{SH}$, i.e. reproduces the reference results with less than 1.2\% error.  Run 2 used a smaller timestep $\Delta t = 0.05 \ t_0$, $N_t = 600$, but the ratio obtained was the same as in Run 1. Run 3, on the other hand, tested the velocity resolution dependence by setting $\Delta v = 0.05 \ v_{th}$, and the obtained heat flux was $q = (0.997488 \pm 1\times 10^{-6})q_{SH}$, reducing the relative error below 0.3\%. At this point we believe the relative error is smaller than the precision of the reference SH flux value, and should thus be taken with a grain of salt.

The next set of small perturbation tests aimed to compare the results of kinetic simulations to those performed using the fluid mode of SOL-KiT. For this purpose, since the fluid model assumes the reference SH heat flux described above from the very start of the simulation, we initialized $f_1$ in the kinetic simulation (Run 4) to a local solution which would give the same heat flux as the one in the fluid run (Run 5). Other than this, the parameters of Run 4 were identical to those of Run 1, and the final heat flux ratio was the same as the one there. 
Run 5 was performed with the same parameters as Run 4, but instead of solving the kinetic equation for the electrons, fluid equations for both electrons and ions were solved. The total changes in the temperature value at its maximum for both runs were compared. It was found that the relative diference in the changes was less than 0.3\%, showing good agreement between the kinetic and fluid model in the small perturbation test.

The final heat conduction test run (Run 6) was done for the fluid model, where the plasma was initialized on a logarithmic spatial grid ($\Delta x = 8.5 \ x_0$ and $\Delta x_L = 0.5 \ x_0$) with the Two-Point Model\cite{Stangeby2000} profile

\begin{equation*}
T(x) = \left(T_{u}^{7/2} + \frac{x}{L}(T_{u}^{7/2}-T_d^{7/2}\right)^{2/7}, \quad n(x) = n_u T_u / T(x),
\end{equation*}
where $n_u=0.368 \times 10^{19}m^{-3}$ and $T_u=18 \text{eV}$ are the upstream density and temperature, while $T_d = 5 \text{eV}$ is the downstream temperature. Boundary conditions were fixed. We expect that this profile should not evolve, and after 30000 collision times (for a reference plasma of $n_0 = 2.5\times 10^{19}m^{-3}$ and $T_0= 10 \text{eV}$), the largest deviation from the initial profile was 0.12 \%, which is due to linear interpolation close to the downstream boundary, with most points having less than 0.01\% deviation. 

\begin{figure*}[h]
\centering
\includegraphics[width=0.5\textwidth]{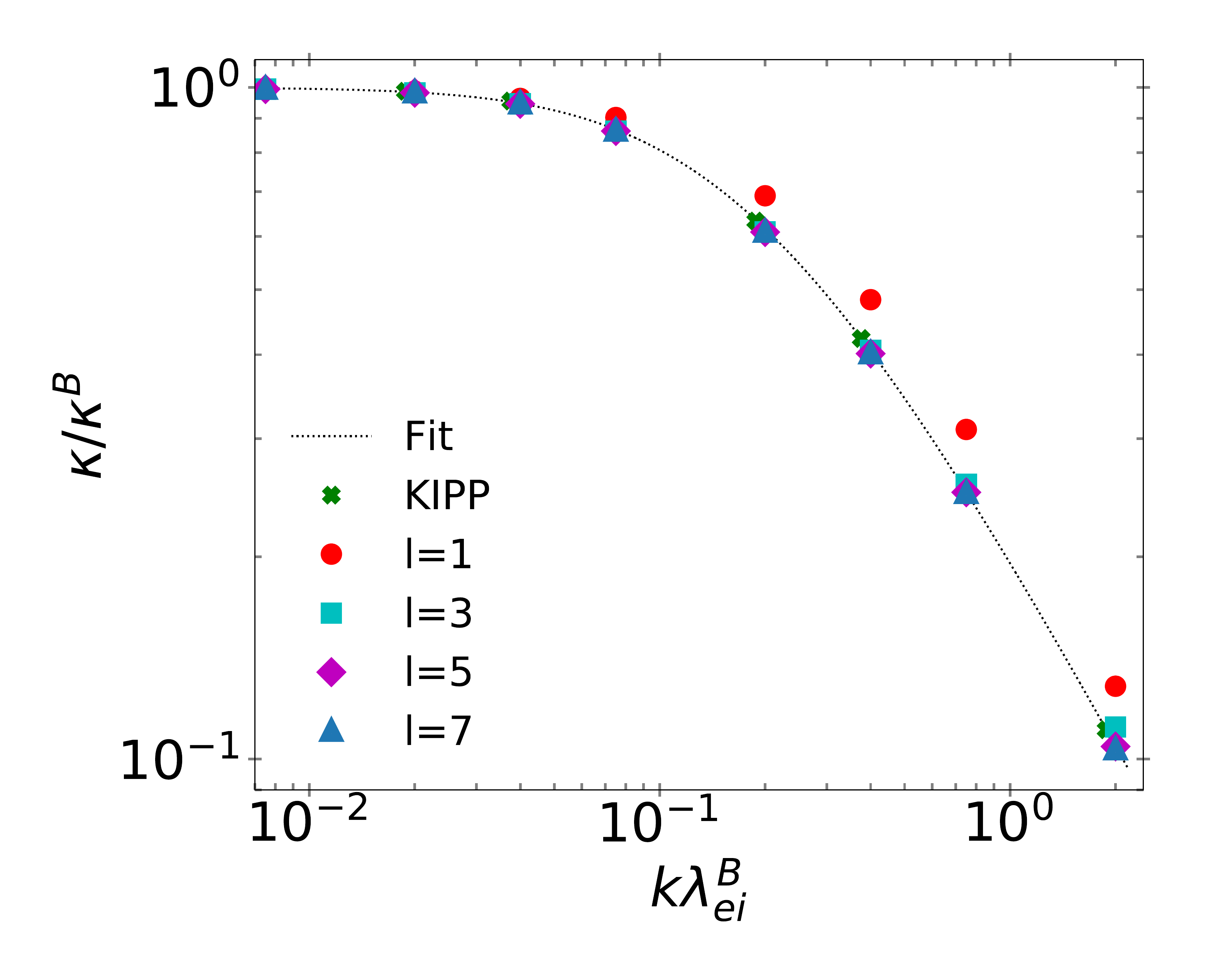}
\caption{Comparison of SOL-KiT and KIPP results for the non-local value of heat conductivity for different number of resolved harmonics. KIPP results have been fitted with a version of the fitting function presented in \cite{Brodrick2017}, here see equation \ref{eq:fit}.}
\label{fig:KIPP_comp}
\end{figure*}

\paragraph{Non-local heat conduction tests}In order to test how SOL-KiT handles a more non-local regime, we have performed a set of runs (Set 1) with a setup similar to Run 1 and related runs, featuring a sinusoidal temperature perturbation. For completeness, we give the common run parameters for Set 1

$$ T_0 = 100 \text{eV},\quad T_1 = 0.1 \text{eV}, \quad n_0 = 10^{19}m^{-3},$$
$$ N_c = 63, \quad N_v = 120, \quad \Delta v_1 = 0.0307, \quad c_v = 1.01.$$
The various runs in the set were performed with different system lengths and with varying number of resolved harmonics $l_{max} = 1,3,5,7$. In Figure \ref{fig:KIPP_comp} we plot the $\kappa/\kappa^{(B)} = q/q_{SH}$ for each of the runs, comparing it with values obtained with the code KIPP\cite{Chankin2015,Zhao2017,Chankin2018} and reported by Brodrick et al.\cite{Brodrick2017}. The ratio $q/q_{SH}$ was obtained by running each run in the set until the ratio equilibriated. The different values in Figure \ref{fig:KIPP_comp} are plotted as a function of $k\lambda_{ei}^{(B)}$, where $\lambda_{ei}^{(B)}= 3(\pi/2)^{1/2}x_0/4$ is the Braginskii electron-ion collisional mean freepath, as used by Brodrick et al., and $k=2\pi/L$ is the perturbation wavenumber. KIPP results were fitted using the following function based on equation (25) in \cite{Brodrick2017} 

\begin{equation}
\frac{\kappa}{\kappa^{(B)}}=\left[1+\left(\frac{1}{a(k\lambda_{ei}^{(B)})^2}+\frac{1}{bk\lambda_{ei}^{(B)}}\right)^{-1}\right]^{-1},
\label{eq:fit}
\end{equation}
where the values obtained for the parameters $a$ and $b$ are $51.409789$ and $4.4682314$, respectively. We show that the increase in number of harmonics leads to an increase in agreement between SOL-KiT and KIPP. Already at $l=5$ SOL-KiT results appear to be very close to the fit values, with the increase to $l=7$ having a negligible effect on the result. We also note that the diffusive approximation ($l=1$) appears to break down around $k\lambda_{ei}^{(B)}=0.075$, while $l=3$ seems to hold up further into the non-local regime. 

\subsection{Collisional-Radiative model and inelastic collision tests}

\begin{figure*}[t]
    \centering
    \begin{subfigure}[t]{0.4\textwidth}
        \centering
        \includegraphics[width = 1.0\textwidth]{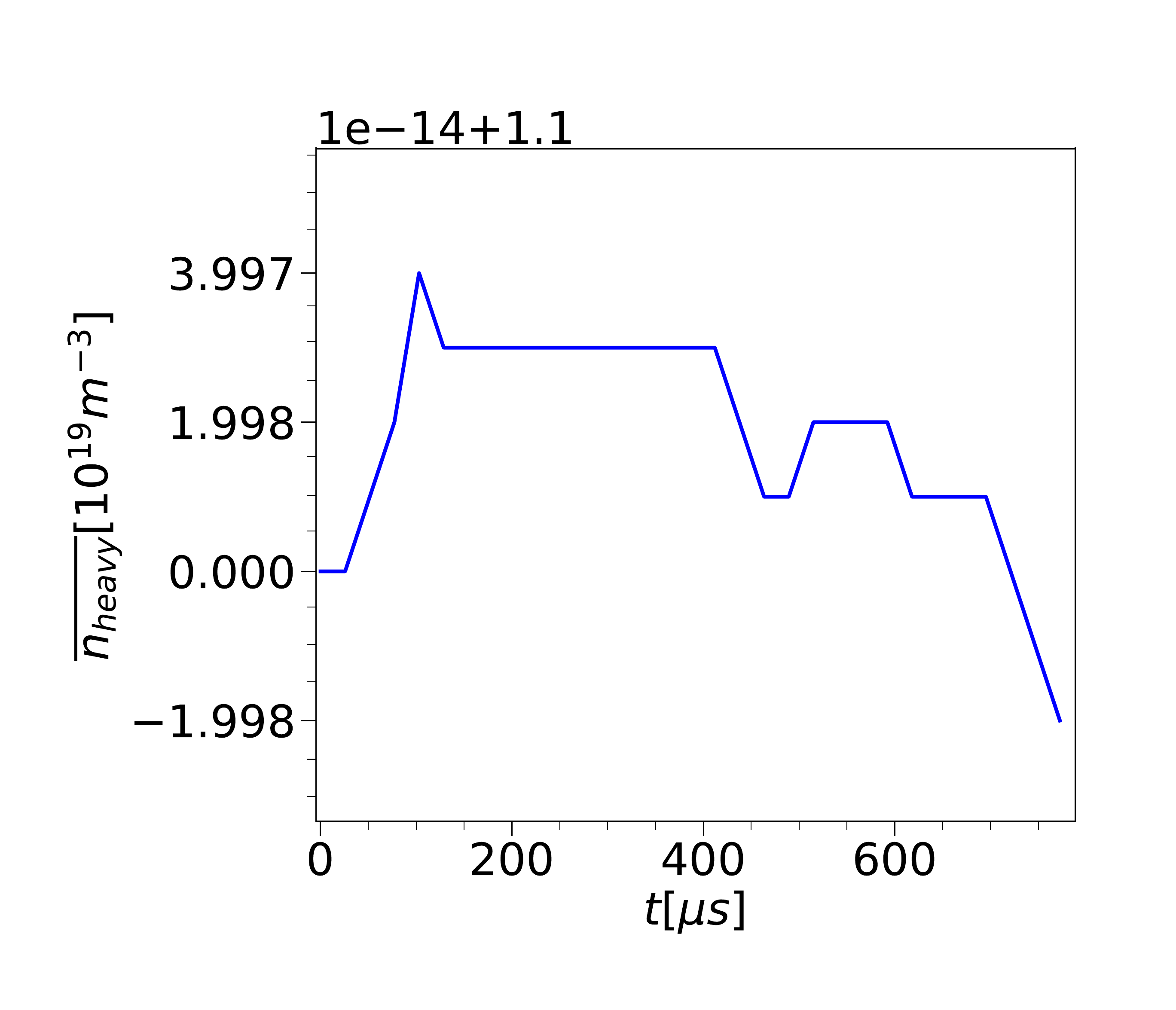}
        \caption{Uniform grid - Run 7}
    \end{subfigure}%
    ~ 
    \begin{subfigure}[t]{0.4\textwidth}
        \centering
        \includegraphics[width = 1.0\textwidth]{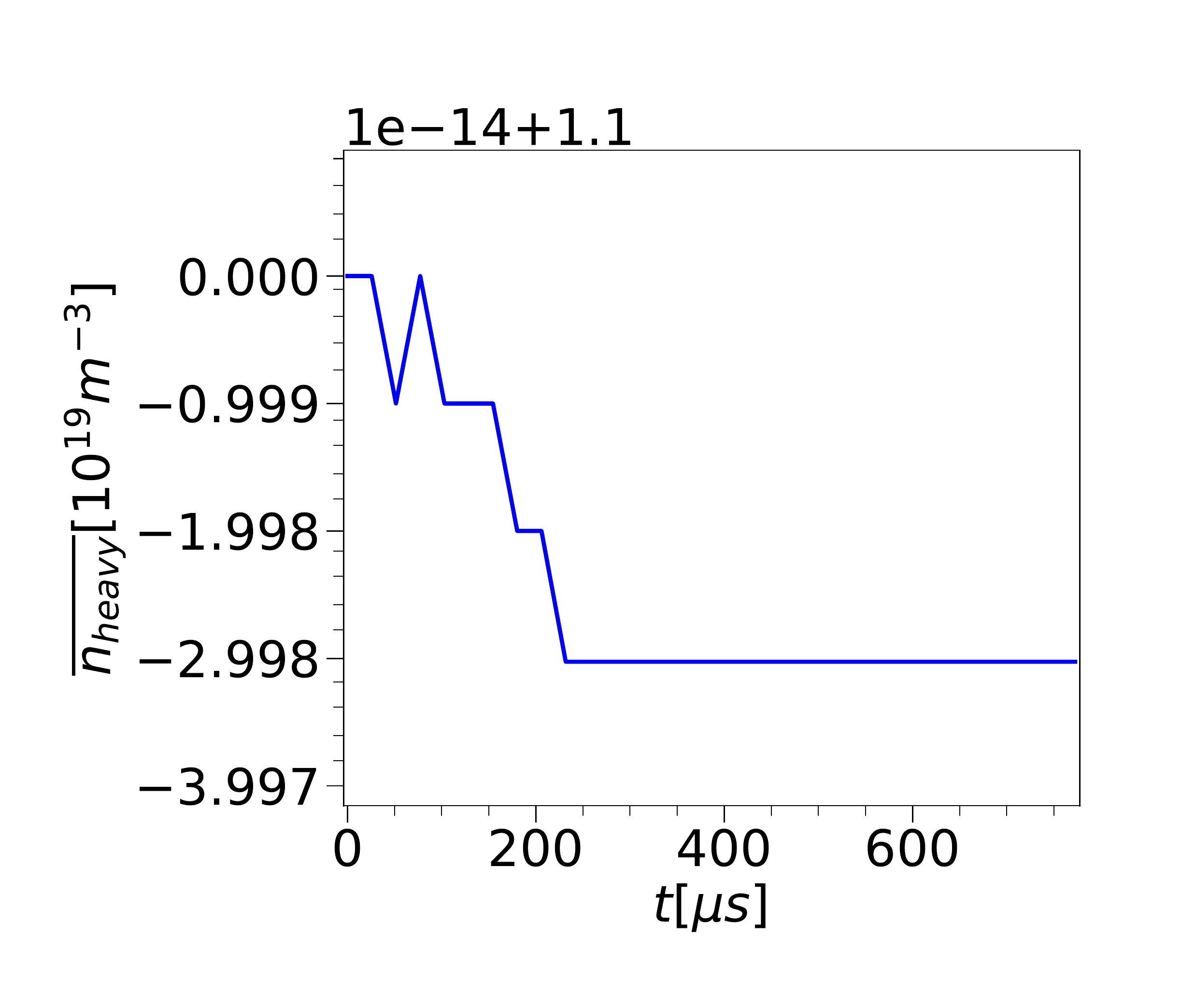}
        \caption{Geometric grid - Run 8}
    \end{subfigure}
    \caption{Evolution of total density in the two inelastic collision discretization test runs.}
    \label{fig:part_cons}
\end{figure*}

The discretization scheme for inelastic collisions presented in this paper is designed with three properties in mind - particle and energy conservation, as well as numerically consistent detailed balance. In this section we present several tests aimed at confirming the desired properties. 

\paragraph{Conservation property tests}We start with two runs differing only in the utilized velocity space grid. The common run parameters are

$$ n_0 =10^{19}m^{-3}, \quad n_e = n_0, \quad n_1 = 0.1 n_0 , \quad N_n = 30,$$
$$\Delta t = 0.5 t_0, \quad N_t =30000, \quad N_v = 120, \quad T_0 = 5 \text{eV}, \quad T_e = T_0,$$
where $N_n$ is the total number of resolved hydrogen states. Both runs included only the inelastic electron-atom processes, and were performed in quasi-0D (low number of spatial cells). The first of the two runs (Run 7) used a uniform grid with $\Delta v = 0.05 v_{th}$, while the second (Run 8) used a geometric grid with $\Delta v_1 = 0.01$ and $c_v = 1.025$. 

\begin{figure*}[h]
\centering
\includegraphics[width=0.5\textwidth]{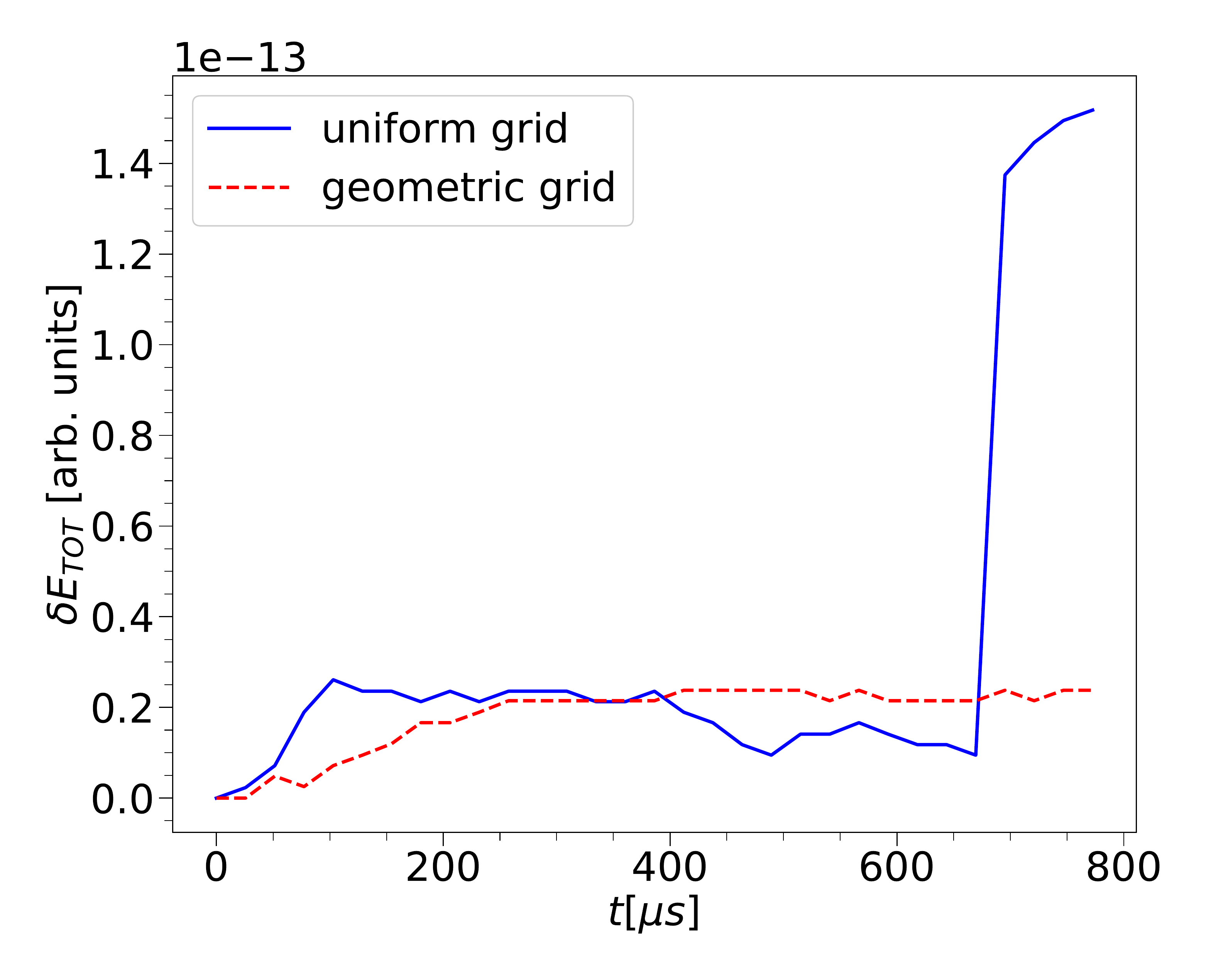}
\caption{Relative change in total energy density in both electron motion and atomic states for the two inelastic collision discretization test runs (Runs 7-8).}
\label{fig:en_cons}
\end{figure*}

Figure \ref{fig:part_cons} shows the average heavy particle density ($n_i + \sum n_b$) in the two runs. As can be seen the total error is of the order of $10^{-14}$, which is consistent with round-off errors. Similar results are obtained for the relative deviation of the total energy density $E_{TOT} = 3n_ekT_e/2 + n_e\epsilon_{ion} +\sum n_b \epsilon_b$ from initial value, where $\epsilon_b$ is the $1 \rightarrow b$ transition energy. This result is shown in Figure \ref{fig:en_cons}, where we see that the geometric grid (Run 8) is performing somewhat better in the later part of the simulation, likely due to a finer velocity grid near the origin.

\paragraph{Detailed balance test}In order to test the numerical detailed balance condition for the inverse processes we treat the electrons as a fluid, thus forcing the distribution function to a Maxwellian. This simulation (Run 9) was performed with the same grid, normalization, and initial conditions as the uniform grid run discussed above. The only difference was the timestep parameters being $N_t = 1000$ and $\Delta t = 100t_0$. By the end of the simulation the atomic states settled close to a Saha-Boltzmann equilibrium, as would be expected from the detailed balance condition. The ionization degree relative error computed against the analytical solution for hydrogen was $\delta X = 1.134\times 10^{-8}$, while the total density and energy errors at the end of the simulation were $\approx 6\times 10^{-14}$ and $\approx 5\times 10^{-14}$, respectively. 

\begin{figure*}[h!]
    \centering
    \begin{subfigure}[t]{0.4\textwidth}
        \centering
        \includegraphics[width = 1.0\textwidth]{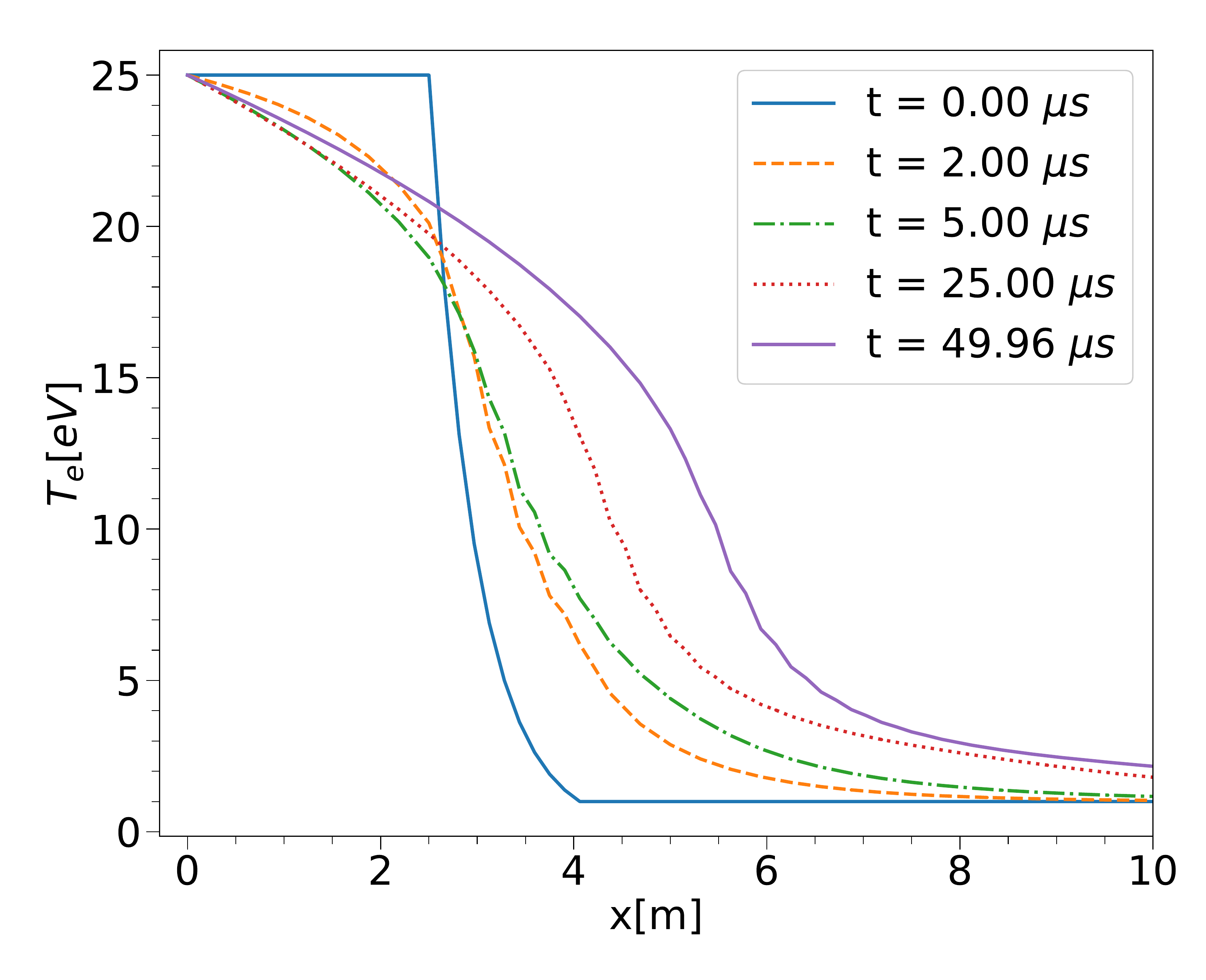}
        \caption{Electron temperature}
    \end{subfigure}%
    ~ 
    \begin{subfigure}[t]{0.4\textwidth}
        \centering
        \includegraphics[width = 1.0\textwidth]{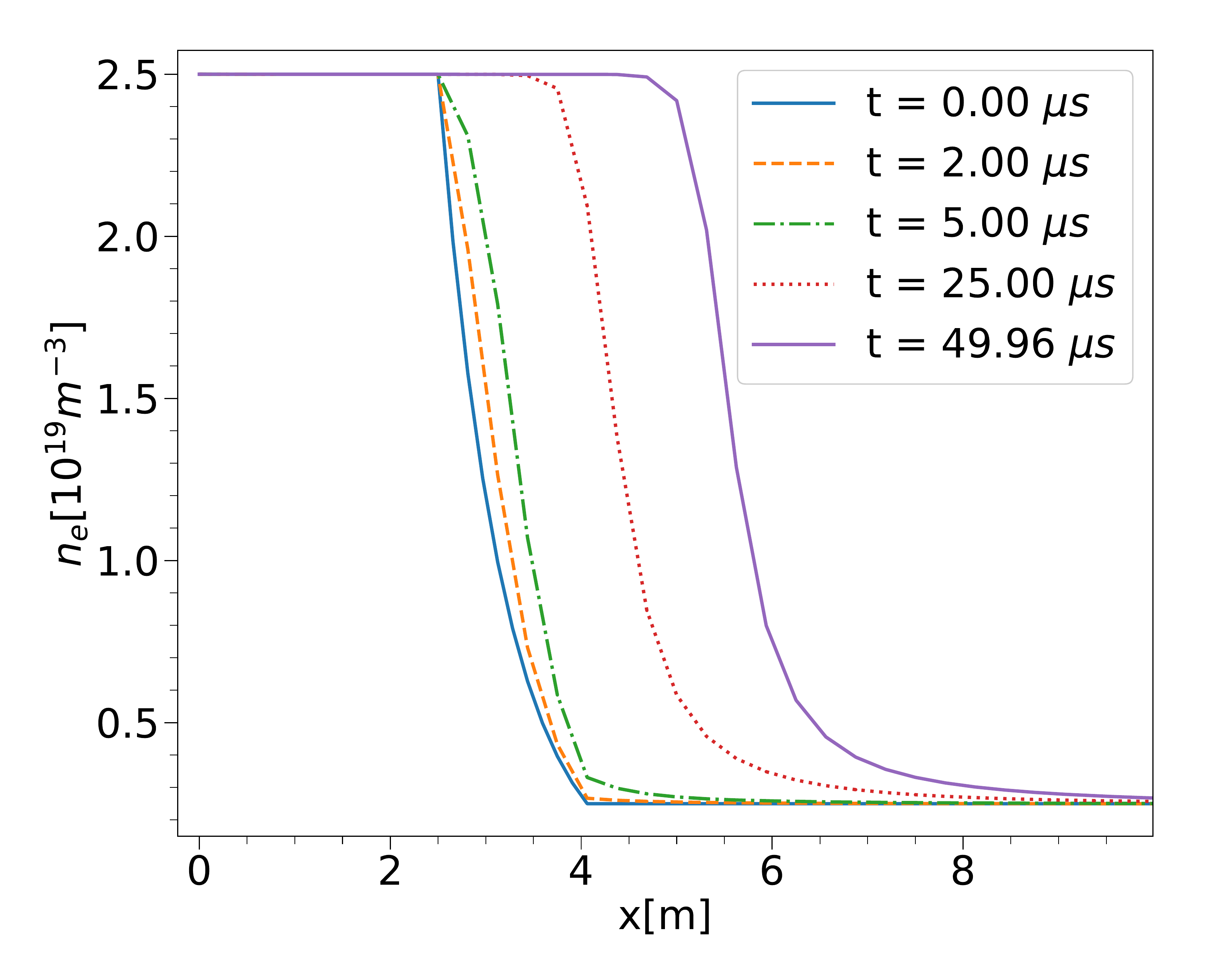}
        \caption{Electron density}
    \end{subfigure}
    \caption{Evolution of temperature and density in reproduction of results from Allais et al. \cite{Allais2005} (Run 10) - corresponds to Figures 1 and 2 from original paper.}
    \label{fig:AM}
\end{figure*}

Note that in all three runs discussed above we disabled all radiative processes, as those would mask any energy conservation processes in the first two runs, and would not allow for equilibrum in the detailed balance test.

\paragraph{Integrated test}Finally, an integrated test (Run 10) was performed in order to test the full interplay of electon and neutral processes. We have attempted to replicate as closely as possible the simulation performed by Allais et al.\cite{Allais2005} using the code FPI. Parameters used in this run were

$$T_0 = 10 \text{eV}, \quad  n_0 =2.5\times10^{19}m^{-3}, \quad N_n = 30, \quad n_{TOT} = n_0,$$
$$N_v = 80, \quad \Delta v_1 = 0.05 v_{th}, \quad c_v = 1.025, \quad l_{max} = 1$$
$$N_c = 64, \quad \Delta x = 5.536x_0, \quad N_t = 16600, \quad \Delta t = 0.1 t_0,$$
amounting to a domain of length $L = 20.31 \text{m}$ and total simulation time $t_{TOT} = 49.96 \mu\text{s}$. At $x=0$ the temperature is initialized to 25 eV, and the plasma is 100\% ionized. From $x=2.5$m to $x=4.06$m the temperature drops exponentially to 1eV, while the ionization degree drops to 10\%. All inelastic collisions were enabled, as well as radiative processes, while ions and neutrals were left stationary as in the original paper. Boundary conditions were fixed to initial values. Figure \ref{fig:AM} shows the evolution of electron temperature and density corresponding to Fig. 1. and Fig. 2. in \cite{Allais2005}. Qualitative behaviour is recovered, while discrepancies of less than 10\% are most likely caused by potential differences in the intialization, as well as likely different spatial and velocity space resolutions (not reported in \cite{Allais2005}). Another potential cause of discrepency is the use different databases in SOL-KiT compared to FPI.

\subsection{Ion flow tests}

\begin{figure*}
    \centering
    \begin{subfigure}[b]{0.35\textwidth}
        \centering
        \includegraphics[width = 1.0\linewidth]{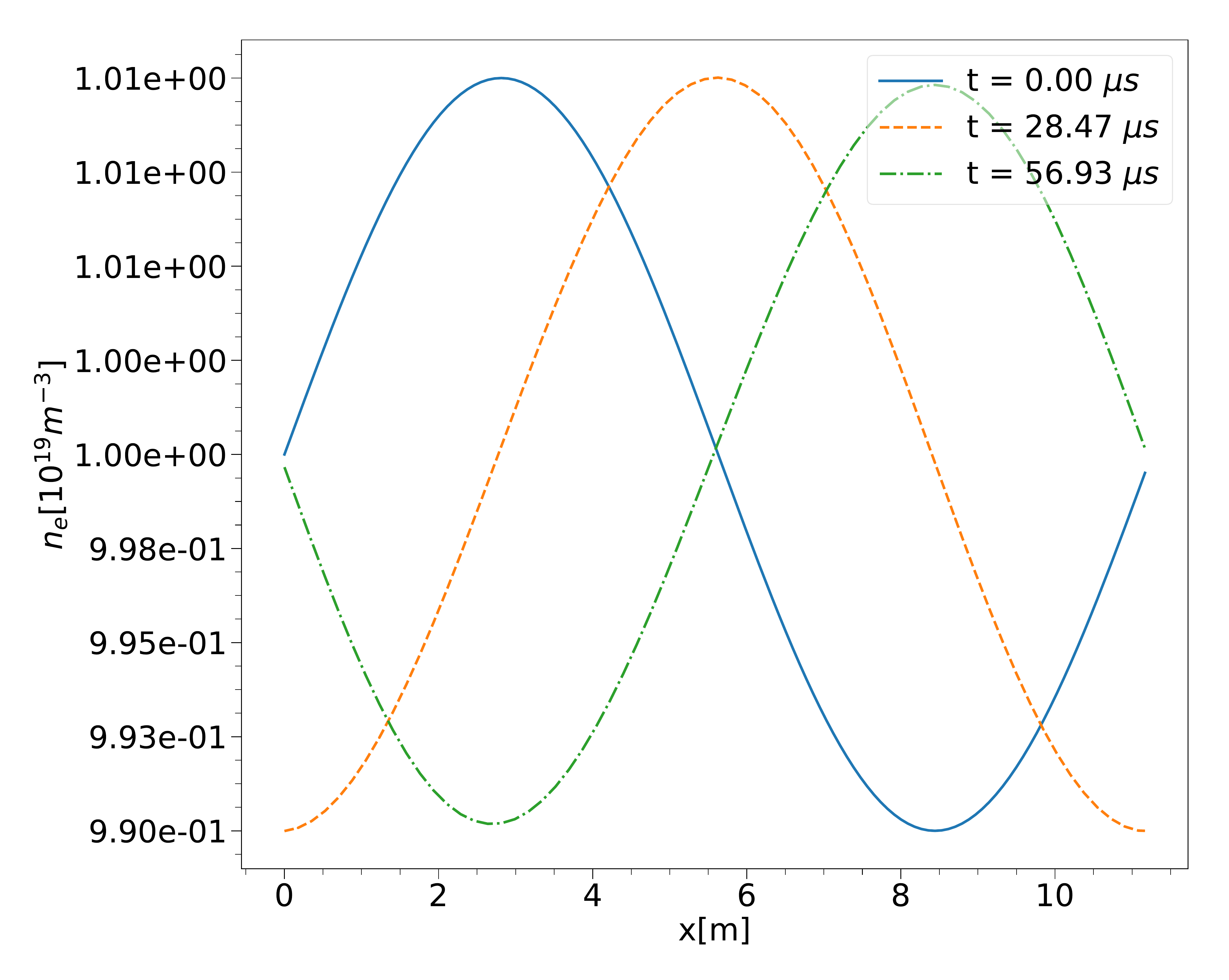}
        \caption{Electron/ion density}
    \end{subfigure}%    
~   
    \begin{subfigure}[b]{0.35\textwidth}
        \centering
        \includegraphics[width = 1.0\linewidth]{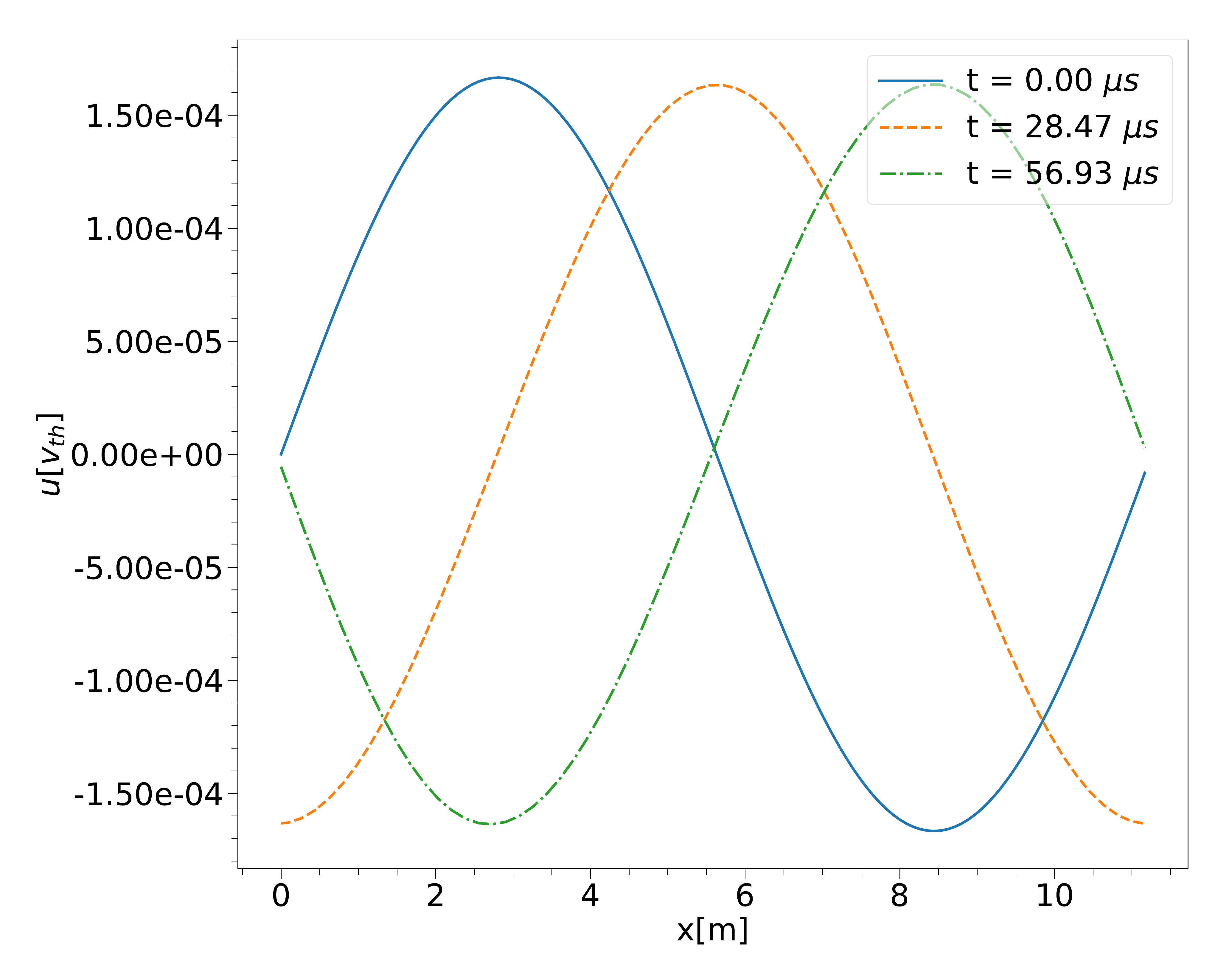}
        \caption{Ion velocity}
    \end{subfigure}
    \caption{Evolution of density and ion velocity during the acoustic wave propagation test from Run 11.}
    \label{fig:acoustic}
\end{figure*}

\paragraph{Acoustic wave test}To test the ion advection, we performed the following isothermal ion acoustic wave test (Run 11). Parameters of the simulation were

$$T_0 = 100 \text{eV}, \quad  n_0 =10^{19}m^{-3}, \quad N_t = 6000,\quad  dt = 0.01t_0, \quad l_{max} = 1$$
$$N_v = 120, \quad  \Delta v = 0.1v_{th},  \quad N_c = 128,\quad  dx = 0.0078125x_0,$$
with the density and ion velocity initialized on a periodic grid of length $L = x_0$ as 

$$n(x) = n_0 + 0.01n_0 \sin\left(2\pi\frac{x}{L}\right), \quad u_i(x) = \frac{v_{th}}{6000}\sin\left(2\pi\frac{x}{L}\right).$$
Electron-electron collisions for $l=0$ and electron-ion collisions were turned on in the simulation. The density and ion velocity are presented at several points during the evolution of the acoustic wave in Figure \ref{fig:acoustic}. The sound speed was evaluated by fitting a sine function to the ion velocity profiles at different times, giving the value $c_s = (1.001 \pm 0.013)c_s^{analytical}$. Estimation of the error was performed conservatively, by taking the greatest deviation from the computed mean sound speed. Nonetheless, the obtained agreement is satisfactory even with the conservative error estimate. 

\begin{figure*}[h]
\centering
\includegraphics[width=0.45\textwidth]{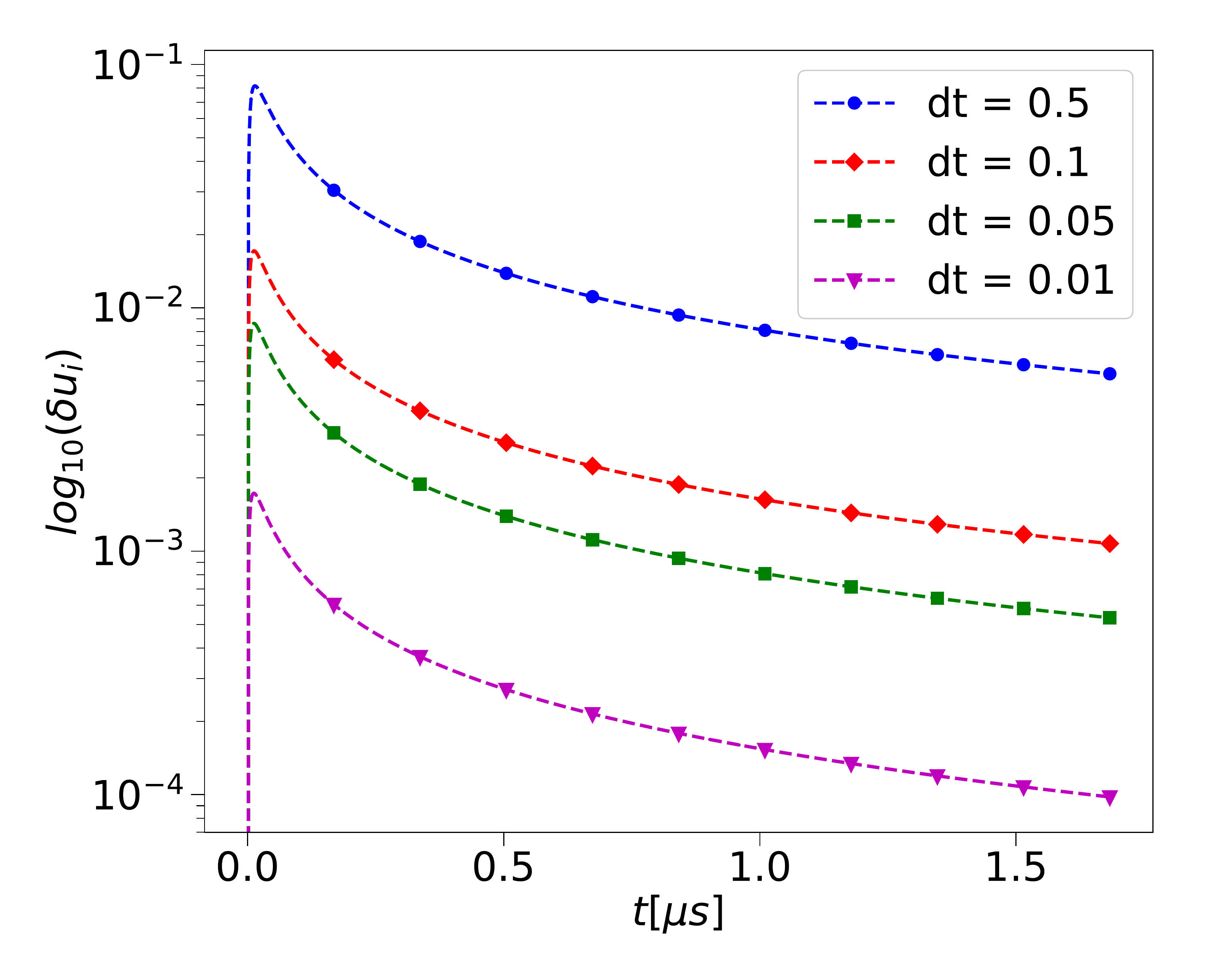}
\caption{Relative difference between analytical and computed values during charge-exchange friction test for various timestep lengths - Set 2.}
\label{fig:CX_test}
\end{figure*}

\paragraph{Charge-exchange friction test} To test the charge-exchange cross-section implementation, a set of runs (Set 2) with the following parameters was performed

$$T_0 = 10 \text{eV}, \quad  n_0 =2.5\times10^{20}m^{-3}, \quad N_n = 1, \quad n_{TOT} = n_0, \quad n_{e,i}=n_0,$$
with the electrons and ions decoupled by turning the $E$-field and electron-ion collisions off. The only operator being tested was the charge-exchange friction term in the ion equation. The total simulation time was kept at 500 e-i reference collision times. The ion velocity was initialized at $u_i = v_{th}$, and its value was compared to the analytical solution $u(t) = u(0)/(1+u(0)n_im_in_1\sigma_{CX,1}t)$ as the ions were slowed down by friction. The relative error of the evolved ion velocity is plotted in Figure \ref{fig:CX_test} for different timestep lengths. It should be noted that the initial value of the ion velocity is unphysically high, and was used only for stress testing the operator, with velocities towards the end of the simulation being more in line with values observed in the SOL.

\subsection{Divertor boundary condition tests}

\begin{figure*}[h]
\centering
\includegraphics[width=0.45\textwidth]{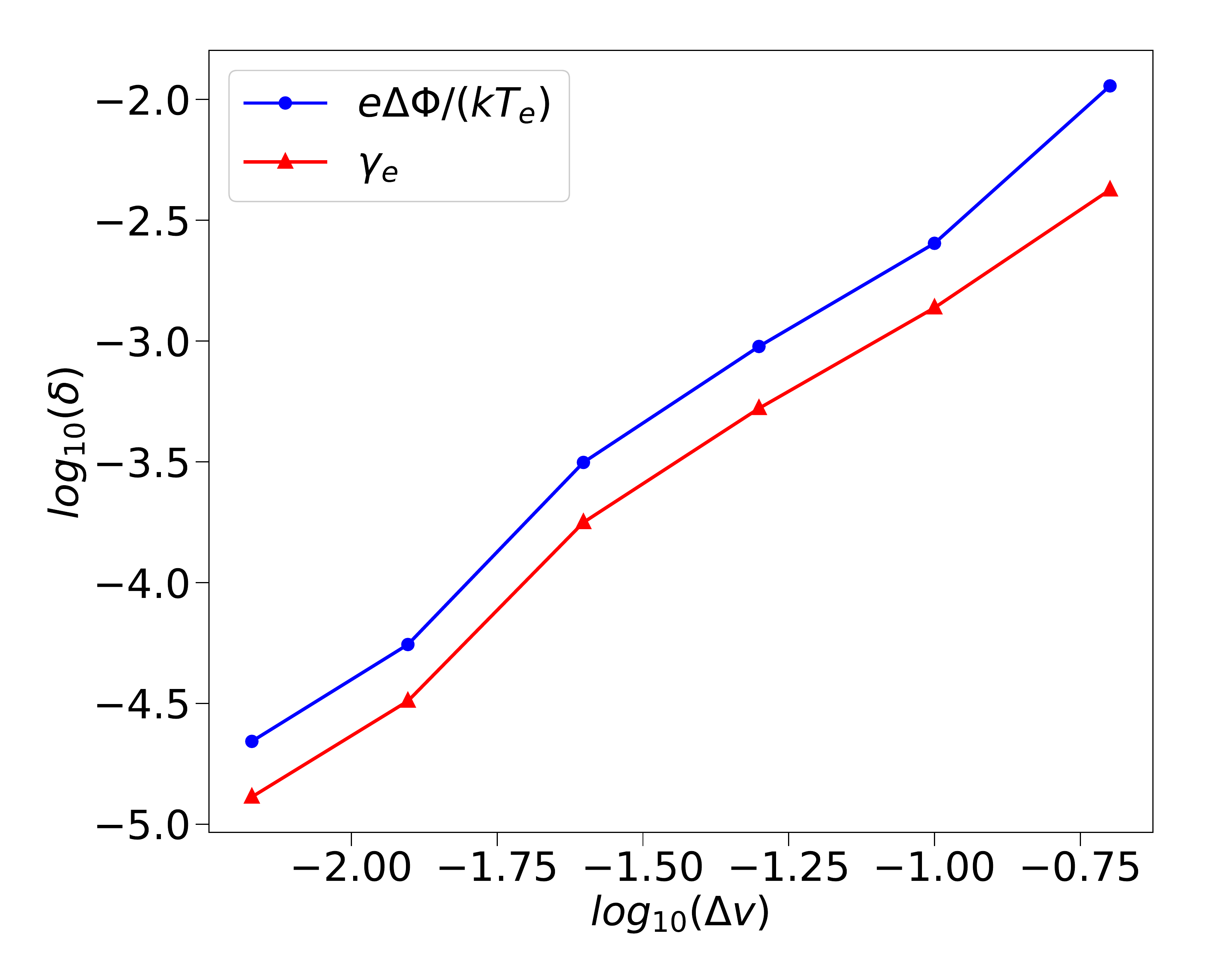}
\caption{Relative error of computed sheath heat transmission coefficient and potential drop as a function of velocity grid resolution - Set 3. Distribution was set to Maxwellian and wasn't evolved in order to compare to the analytical results (see text).}
\label{fig:rel_err_sh}
\end{figure*}

\paragraph{Analytic limit convergence test}The first condition a cut-off logical boundary condition would need to satisfy is the analytical Maxwellian limit for the sheath properties. In order to test this aspect of the operator, the cut-off procedure and calculation of the sheath heat transmission factor and potential drop were performed without evolving the initially Maxwellian distribution. A set of single step $\Delta t =0$ simulations (Set 3) with different velocity space resolutions (constant $v_{max}=6v_{th}$) was used to evaluate convergence of the cut-off calculation without evolving the distribution. The results of this test are shown in Figure \ref{fig:rel_err_sh}, where we see that both the sheath heat transmission coefficient and the potential are within a few percent of the analytical values\cite{Stangeby2000} ( $\gamma_e=2 - 0.5\ln (2\pi(1+T_i/T_e)m_e/m_i)$ and $\Delta \Phi =  \gamma_e - 2$) even for the coarsest grid used.

\begin{figure*}[h]
\centering
\includegraphics[width=0.45\textwidth]{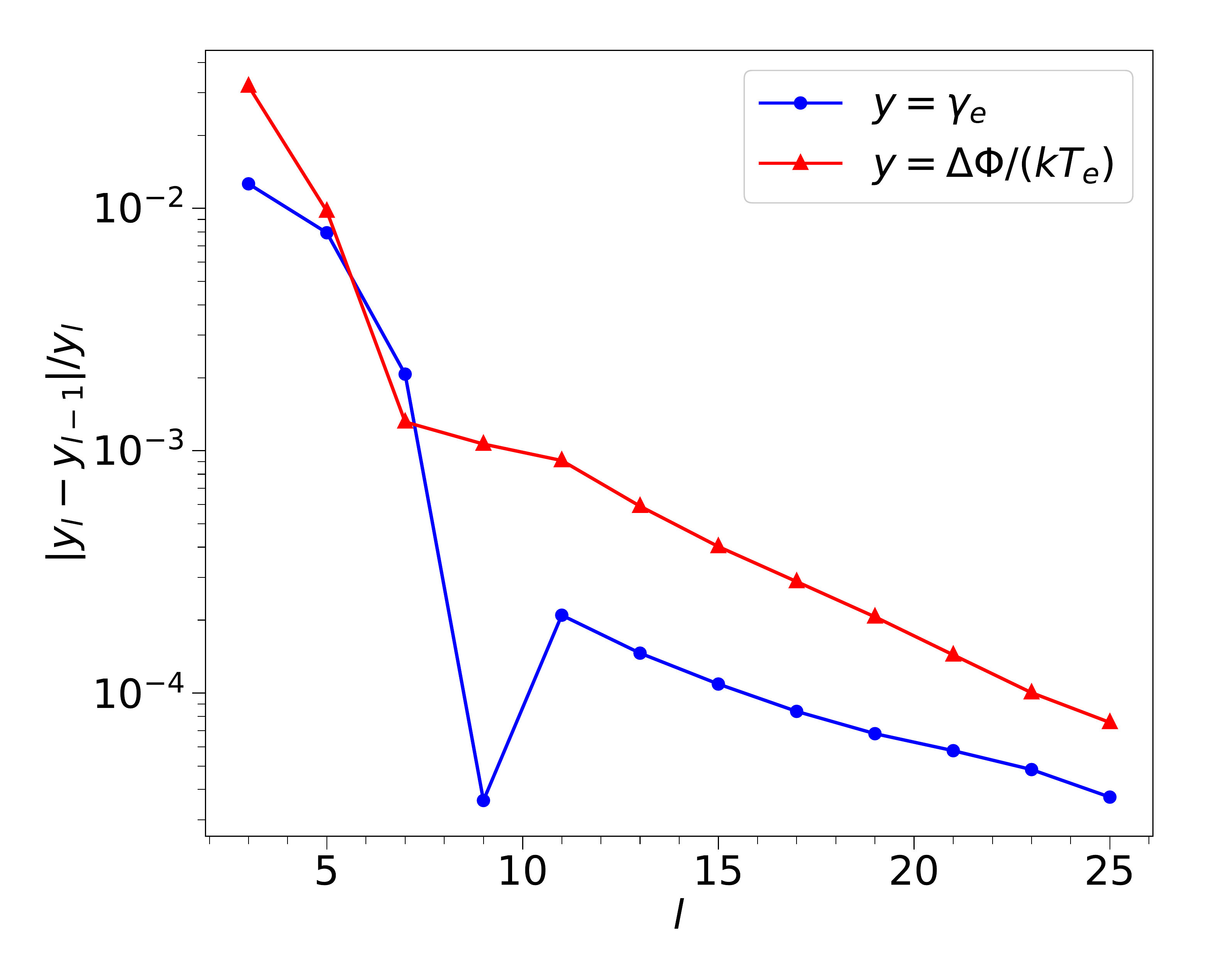}
\caption{Relative change in sheath properties when increasing the number of harmonics in the simulation. Non-monotonic behaviour was observed around $l=9$, causing the drop in relative change - Set 4. }
\label{fig:div_convergence}
\end{figure*}

\paragraph{High harmonic convergence test} In order to test convergence of the sheath properties in a driven system when the number of harmonics is varied a set of runs (Set 4) was performed with the following parameters

$$T_0 = 10 \text{eV}, \quad  n_0 =2.5\times10^{19}m^{-3},$$
$$N_v = 80, \quad \Delta v_1 = 0.05 v_{th}, \quad c_v = 1.01,$$
with a short system $L = 1x_0$ and $N_c=2$. Electron-electron collisions were included only for $l=0$ while electron-ion collisions were on for all $l>0$. A Maxwellian with $T_e=T_0$ and $n_e = n_0$ was imposed as a fixed boundary condition at $x=0$. This provided a scenario where the boundary condition operator was sufficiently stressed without entering a strictly collisionless regime. Simulations were run until the sheath properties equilibriated, and the results of the test are shown in Figure \ref{fig:div_convergence}. Simulations were performed up to $l=25$, and the relative changes in the sheath potential drop and heat transmission coefficient were tracked. Very quickly the change drops below 1\%, before nonmonotonic behaviour is observed around $l=9$. While convergence appears slow, this is to be expected with such a sharply cut-off distribution being expanded in Legendre polynomials. Fortunately, even in this highly stressed scenario a relatively low number of harmonics $l\le5$ seems to be enough to capture the physically important behaviour. 

\section{Discussion}

 In the previous section we have presented tests that show SOL-KiT agreeing with established analytical results for phenomena of interests, namely the heat conduction and logical sheath boundary condition. The Epperlein-Short test (Set 1) presented in Section 4.1. as well as Set 4 in Section 4.4. both show that the code is capable of handling non-local behaviour, as well as demonstrating that suitable convergence in harmonic number $l$ can be achieved. 

The novel discretization scheme for the Boltzman collision integral for inelastic collisions on a velocity grid has been shown to conserve particles and energy up to round-off error. This is to be expected as with $N_n = 30$ hydrogen states, the number of transitions treated, not including radiative transitions, is 930. Given the stiff nature of the Collisional-Radiative model matrix, as well as the corresponding terms in the electron kinetic equations and the large number of transitions, we expected relative errors of order $\approx 10^{-13}$. Runs 7-8 in the previous section agree with this estimate, with the geometric grid notably having better conservation properties. This is likely due to the fact that the high excited state transitions are better resolved with a finer grid near the origin. Detailed balance has also been shown to be observed within a high accuracy, owing to the numerically adapted detailed balance cross-sections derived in Section 3.6.1. 

At the sheath boundary, we implement the logical boundary condition \cite{Procassini1990}. While this is a standard boundary condition in SOL kinetic codes, this is the first time it has been implemented in a harmonic decomposition code. A reasonable concern in this case would be that the sheath properties would be hard to pinpoint with accuracy, given that the sharp cut-off is poorly approximated with Legendre polynomials. While we do notice the effects of the Gibbs phenomenon in the reconstructed distribution function when high $l$ low collisionality systems are treated, we show in Set 4 that a relatively low number of harmonics is sufficient to provide an estimate of the relevant sheath properties. Furthermore, the Maxwellian analytic limit of the sheath properties requires only resolving up to $l=1$. This further supports the notion that, while important in general, high harmonic approximations of the cut-off are not necessary for the transport calculations at the sheath boundary.

SOL-KiT is a fully implicit code, and is thus not limited by the CFL condition. It is, therefore, able to resolve the lowest velocity cells for higher harmonics, unlike explicit codes \cite{Tzoufras2011}. This is one of the key features allowing the implementation of both lab frame ion flow, as well as inelastic electron-neutral collisions, with the former being the main beneficiary of the implicit scheme. The reason for this is that the effect of flowing ions in the lab frame is visible primarily in the low velocity cells, where electrons are highly collisional and are dragged with the ions. A scheme that cannot resolve those cells must resort to calculations in the ion centre of mass frame\cite{Ridgers2008,Epperlein1988}. Given the complicated sheath boundary condition necessary in the simulation of the SOL, the lab frame is the natural choice. The effect of this choice is that the complexity normally arising in various LHS operators (in the Vlasov part) of the electron kinetic equation is transferred into the electron-ion collision operator. In this work the ions are treated as a cold beam for the sake of this collisional computation. While the addition of ion temperature effects would in theory increase the fidelity of the operator, it would strongly depend on the velocity grid resolution around the ion flow velocity. It is likely that the effect would be negligible as the ion thermal velocity is much smaller than the electron thermal velocity, and that the main collisional effect of the ion thermal population is well captured by the delta function of a cold beam. However, refinement of the electron-ion collision operator is a potential path in the development of the code.

Momentum is not explicitly conserved in the electron-electron collision operator for $l>0$ as implemented in the code. As suggested by Tzoufras et al. \cite{Tzoufras2011}, we ensure total momentum conservation by transfering the momentum lost in electron-electron collisions to the ions. This produces a spurious electric field in the regions of high flow speeds. However, this electric field is negligible compared to the physical electric fields occuring in realistic SOL simulations, as the region of non-zero ion flow tends to be in front of the divertor target, where the pre-sheath electric field is generated through pressure gradients. 

The way we calculate the electric field also ensures that quasineutrality is observed up to displacement current effects. This allows SOL-KiT to treat collisionless phenomena on the Debye length scale. However, as the logical sheath boundary condition does not require resolving the Debye sheath, this regime has not been fully explored. 

Performance of the code was tracked in several of the presented runs. Specifications of the used workstation include an Intel® Xeon(R) CPU E5-2630 v3 @ 2.40GHz with a maximum of 16 threads, and 15.6GB of RAM. In all runs performed so far it appears that the CPU requirement always outweighs the memory needs of the code. Execution times on the workstation during tests varied from under 6 minutes for the local heat conductivity runs, to 18.5h for the reproduction in Run 10. Most runs were performed with the full 16 threads, excepts for the several quasi-0D tests. These runs (such as Runs 7-8) took just under 7h to complete on a single core. Run 9, with both electrons and ions treated as a fluid took slightly over 3h. Since fluid runs have both a simpler matrix, as well as no requirement to resolve the collisional times, they tend to run much faster and for longer physical simulation times. The two main bottlenecks in the current version of the code have been identified as the electron-electron collisions for $l>0$ as well as the electron-neutral collision operators. As was noted above, the e-e collisions for $l>0$ produce dense submatrices for each harmonic, both increasing the matrix computation times as well as increasing the difficulty of solving the matrix system. The e-n collisions similarly produce almost dense matrices, but the main bottleneck there is the large number of transitions, translating into hundreds of effective collision operator computations per timestep. In order to speed this up, the detailed balance cross-sections can be updated only once per timestep, as opposed to during every nonlinear iteration. Another way of speeding the neutral physics up would be grouping the higher excited states into effective states, thus reducing the total number of collision integral matrices computed every timestep. 

While SOL-KiT is implicit, in practice the timestep is limited by the capabilities of the solver. We currently use the Bi-CGSTAB iterative matrix solver with Block Jacobi preconditioning in the PETSc package. This solver tends to struggle when the timestep is many ($\approx50$) times the electron-ion collisional time, as well as when higher harmonics are included due to the added stiffness of the matrix due to the fast evolution of $f_l$ for high $l$'s. However, in most situations we are aiming to resolve the collisional phenomena, and the timestep-limiting effect of higher harmonics becomes evident at a number of harmonics already high enough where the dense matrix effects already cause the majority of the slowdown.

SOL-KiT is currently parallelized using MPI, with domain decomposition along the $x$-axis. Basic optimization was performed on the code, however, many avenues of potential improvement remain. Besides the already mentioned grouping of excited neutral states, parallelization in $l$ number is also being considered, though the degree of speedup attainable would depend heavily on the inter-processor communication. Different preconditioners and solvers could also be employed depending on the problem, though this would require more in-depth tests of the code's performance. 

One of the code's design goals was to enable consistent and convenient one-to-one comparisons between a fluid and kinetic model of electron transport. This was accomplished by ensuring that the physics implementated in the kinetic and fluid modes use the same data (e.g. cross-sections etc.). A possible extension of this approach would be the inclusion of various non-local fluid models as comparison options. The easiest to implement would possibly be the flux limiter method \cite{Fundamenski2005}, and more complicated models like the SNB model and others\cite{Brodrick2017} could be implemented self-consistently with the current framework. 

The elastic electron-neutral collision operator currently implemented assumes a simple cross-section (see 2.1.3.). As we believe the cross-section could be improved with a more detailed model based on experimental data, while tested, this operator is not in regular use. Improvements of the neutral model are being planned, including potential additions of molecules, the implementation of a fluid as opposed to a diffusive neutral model, and the option to include high $Z$ impurities. To accompany the improved neutral model, the next step in SOL-KiT's development will be the addition of an ion temperature equation, as well as an electron-ion collision operator for $l=0$. This will allow us to probe more varied SOL regimes, where the ion and electron temperatures are decoupled. 

Compared to PIC codes, the finite difference approach taken with SOL-KiT provides a speed-up in computation, as well as avoiding the noise issues present in PIC codes. The closest comparison to SOL-KiT, however, is the code FPI \cite{Abou-Assaleh1992,Abou-Assaleh1994,Allais2005}. While a form of the logical boundary condition appears to have been implemented in FPI  \cite{Abou-Assaleh1992,Abou-Assaleh1994}, to our knowledge no in-depth discussion of its Legendre formalism form is available. Similarly, electron-neutral inelastic collisions have been implemented and reported in both ALLA\cite{Batishchev1999} and FPI, but the conservative properties of the operators were not explored in detail. In this paper we report in detail both of these numerical aspects in the context of a Legendre decomposition model. Furthermore, the combination of a fully implicit algorithm together with the lab frame ion treatment and the aforementioned numerical facets has been realized for the first time in SOL-KiT. The addition of a self-consistent fluid mode for the electrons adds another aspect to the code, with work on coupling the two modes under way. 

\section{Conclusion}

 We have  presented the model and the numerics behind the newly developed fully implicit arbitrary harmonic electron kinetic code SOL-KiT. The code is designed for the exploration of kinetic effects in the Scrape-Off Layer, and includes a self-consistent fluid electron mode for one-to-one comparisons. Novel conservative implemenentation of the electron-neutral collision operators for inelastic collisions and the Legendre polynomial version of the logical sheath boundary condition are presented and tested, showing good agreement with expected properties. We have shown that the code can resolve highly non-local effects utilizing high harmonics, and have demonstrated that some of the more demanding operators converge well with the increase in the number of harmonics. The next steps in SOL-KiT development have been laid out, focusing on both improving performance, as well as adding new physics to extend the applicability of the code.

\section*{Acknowledgements}

This work was supported through the Imperial President's PhD Scholarship scheme. This work was partially funded by the RCUK Energy Programme [grant number: EP/P012450/1]. Some of the simulation results were obtained with the use of the Imperial College Research Computing Service \cite{CX1}

\appendix 

\section{Divertor target boundary condition transformation matrix derivation}

\begin{figure}
\centering
\includegraphics[width=0.5\textwidth]{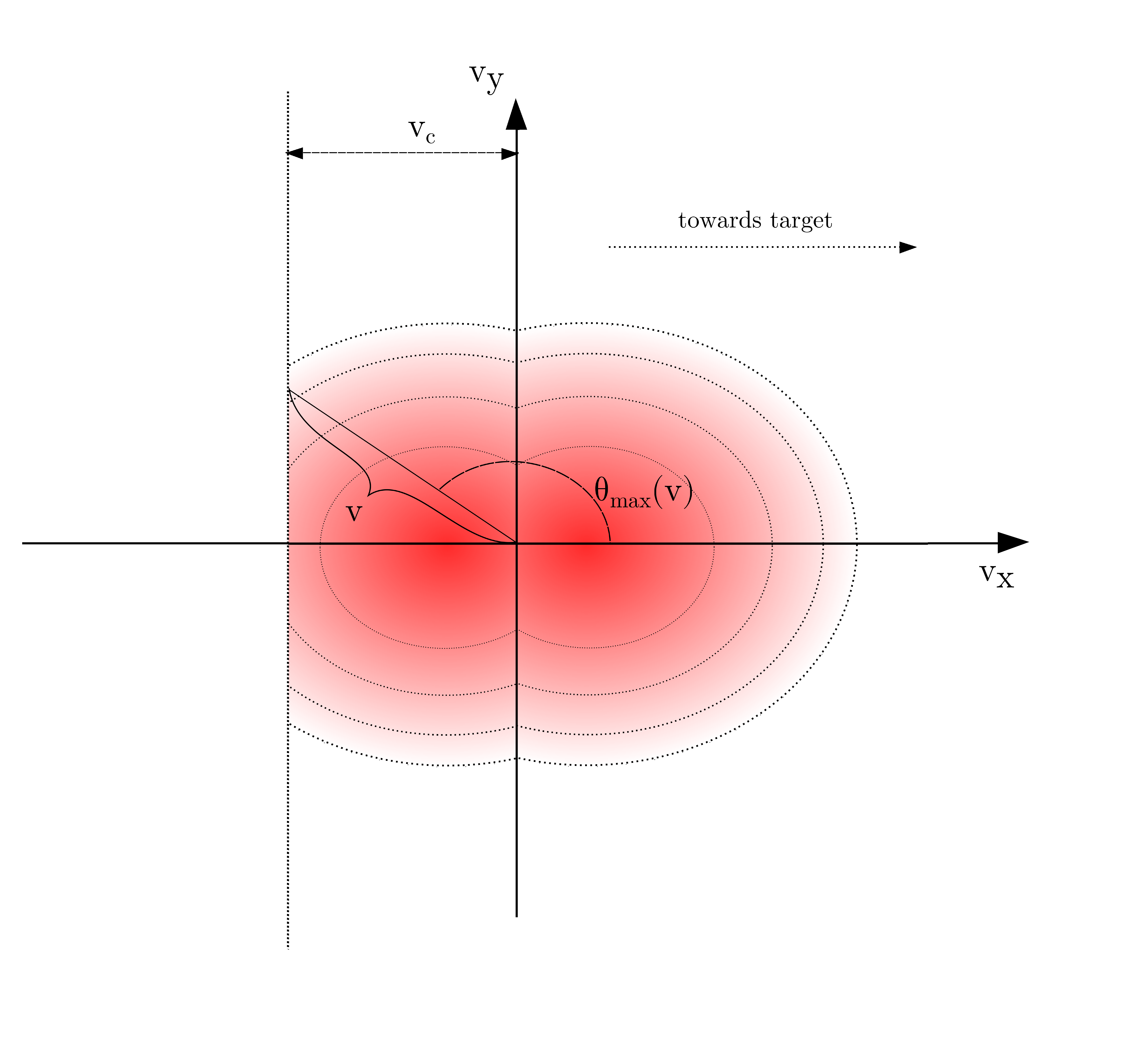}
\caption{The distribution function (anisotropy exaggerated) after applying reflection and cut-off. Highlighted is the integration limit $\theta_{max}(v).$}
\label{fig:cut-off}
\end{figure}

As stated above, we label the cut-off velocity $v_c$ and take that all electrons with parallel velocity greater than $v_c$ are lost to the sheath, and all other electrons are reflected. This informs the form of the distribution function at the sheath entrance in the following way. If we know the distribution function for positive $v_x$ (moving towards the divertor target) and can expand it into Legendre polynomials (assuming azimuthal symmetry, of course)

$$f(v_x>0,v_\perp) = \sum_l f_l(v) P_l(\cos \theta), \quad \cos \theta > 0.$$
Now, if we were to reflect all electrons back, the total (reflection included) distribution function would be

 $$f_R(v, \theta) = \sum_l f_l(v) P_l(|\cos \theta |).$$
Finally, we take away all of the electrons that would have been lost to the sheath and thus should not be in the distribution function. This is simply all electrons with $v \cos \theta=v_x<-v_c$. Using a Heaviside step function, this can be expressed as

\begin{equation}
f_c(v, \theta) = f_R(v,\theta) \Theta (v \cos\theta+v_c).
\label{eq:fc}
\end{equation}
From here we can use Legendre polynomial orthogonality to extract the $l$-th harmonic of the ``cut-off'' distribution

 \begin{equation}
f_{cl}(v) = \frac{2l + 1}{2} \int_{-1}^{1} f_R(v,\theta) \Theta (v \cos\theta+v_c) P_l(\cos\theta) d(\cos\theta).
\label{eq:fcl}
\end{equation}
Using the decomposition of $f_R$ we write equation (\ref{eq:fcl}) as

 \begin{equation}
f_{cl}(v) = \frac{2l + 1}{2} \sum_{l'} f_{l'}(v)  \int_{-1}^{1}P_{l'}(|\cos \theta |) P_l(\cos\theta)  \Theta (v \cos\theta+v_c)d(\cos\theta).
\label{eq:fcl2}
\end{equation}
Here it's helpful to visualize the two-dimensional velocity space of the cut-off. This is shown in Fig \ref{fig:cut-off}. As can be seen, the integral in (\ref{eq:fcl2}) has different limits depending on the value of the velocity $v$. This can be written as (taking $x=\cos \theta$)

\begin{equation}
f_{cl}(v) = \frac{2l + 1}{2} \sum_{l'} f_{l'}(v)  \int_{x_{min}}^{1}P_{l'}(|x|) P_l(x) dx,
\label{eq:fcl3}
\end{equation}
where $x_{min} = \max(-1,\cos \theta_{max}(v))$, with $\cos \theta_{max} = -v_c/v$. Introducing the transformation matrix $P_{ll'}$ ($v$ dependence implied)

\begin{equation}
P_{ll'} = \frac{2l + 1}{2}  \int_{x_{min}}^{1}P_{l'}(|x|) P_l(x) dx,
\label{eq:Pll}
\end{equation}
we get a compact expression for the sheath edge electron distribution harmonics

\begin{equation}
f_{cl}(v) = \sum_{l'}P_{ll'} f_{l'}(v).
\label{eq:fcl4}
\end{equation}
Now the problem is reduced to computing the matrix $P_{ll'}$. First, let us dispose of the absolute value in the argument of $P_{l'}$ by separating the integral into positive and negative $x$ intervals

\begin{equation}
P_{ll'} = \frac{2l + 1}{2}\left( (-1)^{l'} \int_{x_{min}}^{0}P_{l'}(x) P_l(x) dx +\int_0^{1}P_{l'}(x) P_l(x) dx\right) ,
\label{eq:Pll2}
\end{equation}
where we have used the parity property of Legendre polynomials $P_l(-x)=(-1)^lP_l(x)$. In order to reduce the integrals in (\ref{eq:Pll2}) to forms available in the literature we expand the integration range in the first integral up to $1$ and after grouping terms up, we get

\begin{equation}
P_{ll'} =P^x_{ll'}+P^0_{ll'},
\label{eq:Pll3}
\end{equation}
where

\begin{equation}
P^x_{ll'}=\frac{2l + 1}{2}(-1)^{l'} \int_{x_{min}}^{1}P_{l'}(x) P_l(x) dx,
\label{eq:Pllx}
\end{equation}

\begin{equation}
P^0_{ll'}=\frac{2l + 1}{2}(1-(-1)^{l'}) \int_{0}^{1}P_{l'}(x) P_l(x) dx.
\label{eq:Pll0}
\end{equation}
The integral in (\ref{eq:Pllx}) can be found in the literature (see for example \cite{Szmytkowski2006} for general case) and well known recurrence formulae for the derivatives of Legendre polynomials can be used to get 

\[
  P^x_{ll'}=\begin{cases}
               (-1)^{l'} \delta_{ll'} \quad \text{if $x_{min}  = -1$}\\
               F_{ll'} \quad \text{if $x_{min} > - 1$ and $l \neq l'$}\\
               F_{ll} \quad \text{if $x_{min} > - 1$ and $l = l'$}
            \end{cases}
\]
where

\begin{equation*}
\begin{split}
F_{ll'} &= (-1)^{l'}\frac{2l+1}{2}\Big[-\frac{x_{min}}{l+l'+1}P_l(x_{min})P_{l'}(x_{min}) \\&+\frac{1}{(l-l')(l+l'+1)}\left(lP_{l-1}(x_{min})P_{l'}(x_{min})-l'P_l(x_{min})P_{l'-1}(x_{min})\right)\Big],
\end{split}
\end{equation*}
and 

\[F_{ll} = (-1)^{l}\left[\frac{1}{2} - \left(\frac{x_{min}}{2}[P_l(x_{min})]^2 + \sum_{k=1}^{l-1}P_k(x_{min})(x_{min}P_k(x_{min})-P_{k+1}(x_{min}))\right)\right].\]
For $P^0_{ll'}$ we get 

\[
  P^0_{ll'}=\begin{cases}
               0 \quad \text{if $l'$ even}\\
               \delta_{ll'}\quad \text{if $l',l$ odd and $l = l'$}\\
               0\quad \text{if $l',l$ odd and $l \neq l'$}\\
               \frac{2l+1}{(l-l')(l+l'+1)}\left(lP_{l-1}(0)P_{l'}(0)-l'P_l(0)P_{l'-1}(0)\right) \quad \text{if $l'$ odd and $l$ even}
            \end{cases}
\]
It should be noted that for $v<v_c$ the above equations give $P_{ll'}=0$ for all odd $l$, which is the consequence of reflection symmetry, from which one can also see that there is no effective flux contribution for $v<v_c$ .

\bibliography{Mijin_CPC_Preprint.bbl}

\end{document}